\documentclass[10pt,conference]{./IEEEtran}
\usepackage{tabularx}
\usepackage{multirow}
\usepackage{graphicx}
\usepackage[british]{babel}
\usepackage{cite}
\usepackage{amsmath,amssymb,amsfonts}
\usepackage{algorithmic}
\usepackage{graphicx}
\usepackage{textcomp}
\usepackage{xcolor}
\usepackage{./subcaption}
\usepackage{stmaryrd} %% --RW-- Provides \shortrightarrow amongst other things
\usepackage{flushend} %% --RW-- Balance the last page contents
\usepackage{tikz}
\newcommand*\circled[1]{\tikz[baseline=(char.base)]{
                    \node[shape=circle,draw,inner sep=2pt] (char) {#1};}}

\begin{document}

%
% The "title" command has an optional parameter, allowing the author to define a "short title" to be used in page headers.
\title{Boomerang: Real-Time I/O Meets Legacy Systems}

\author{\IEEEauthorblockN{Ahmad Golchin}
\IEEEauthorblockA{\textit{Computer Science Department} \\
\textit{Boston University}\\
Boston, USA \\
golchin@cs.bu.edu}
\and
\IEEEauthorblockN{Soham Sinha}
\IEEEauthorblockA{\textit{Computer Science Department} \\
\textit{Boston University}\\
Boston, USA \\
soham1@cs.bu.edu}
\and
\IEEEauthorblockN{Richard West}
\IEEEauthorblockA{\textit{Computer Science Department} \\
\textit{Boston University}\\
Boston, USA \\
richwest@cs.bu.edu}
}

\maketitle
\thispagestyle{plain}
\pagestyle{plain}

\begin{abstract}
This paper presents Boomerang, an I/O system that integrates a legacy
non-real-time OS with one that is customized for timing-sensitive
tasks. A relatively small RTOS benefits from the pre-existing
libraries, drivers and services of the legacy system.  Additionally,
timing-critical tasks are isolated from less critical tasks by
securely partitioning machine resources among the separate
OSes. Boomerang guarantees end-to-end processing delays on input data
that requires outputs to be generated within specific time bounds.

We show how to construct composable task pipelines in Boomerang that
combine functionality spanning a custom RTOS and a legacy Linux
system. By dedicating time-critical I/O to the RTOS, we ensure that
complementary services provided by Linux are sufficiently predictable
to meet end-to-end service guarantees. While Boomerang benefits from
spatial isolation, it also outperforms a standalone Linux system using
deadline-based CPU reservations for pipeline tasks. We also show how
Boomerang outperforms a virtualized system called ACRN, designed for
automotive systems.

\end{abstract}


\section{Introduction}
\label{sect:intro}
%%%%%%%%%%%%%%%%%%%

Mixed-criticality systems require the spatial and temporal isolation of tasks
to meet timing, safety and security
constraints~\cite{Leiner:SAFECOMP2007}. Additionally, these systems involve
real-time task pipelines to implement sensing, processing and actuation. For
example, an automotive system supports low-criticality infotainment services,
which must be isolated from highly critical driving assistance tasks that
process sensor data to avoid vehicle collisions. 

Spatial isolation ensures that one software component cannot alter another
component's private code or data, or interfere with the control of its
devices. Temporal isolation ensures that a software component cannot affect
when another component accesses a resource (e.g., a CPU). Lack of temporal and
spatial isolation leads to potential timing or functional failures. Failure of
a highly critical task has potentially catastrophic consequences, while
failure of a low-criticality task has less significant consequences.

One way to support mixed-criticality systems is to partition tasks onto
separate hardware. This ensures less critical tasks are unable to directly
affect those of greater importance. Automotive systems have traditionally
taken this approach, by assigning a different functional component to a
separate electronic control unit (ECU)~\cite{tttech}. However, as the
complexity of these systems increases, hardware costs, wiring and packaging
become prohibitive.  
%% Considering automotive systems once again, it is not
%% uncommon to see vehicles with dozens of ECUs assigned to chassis, body, and
%% powertrain functions~\cite{automotive_sensors}; as these systems adopt
%% increased driver assistance and autonomous technologies, it is necessary to
%% consider new hardware platforms rather than add more ECUs. 
For this reason, new hardware platforms that integrate the functionality of
multiple hardware components, including multicore processors, accelerators,
GPUs, and various input/output (I/O) interfaces are now emerging. Tesla's
AutoPilot 2.x, for example, already uses platforms such as the Nvidia Drive
PX2 in its cars, to assist with vehicle control.

An integrated solution, combining tasks of different criticality levels on the
same hardware, requires an operating system to correctly enforce temporal and
spatial isolation. Partitioning operating systems such as
Tresos \cite{Leiner:SAFECOMP2007} and LynxOS \cite{lynx} have been developed
for automotive and avionics systems, respectively, in accordance with
standards such as AU\-TOSAR \cite{autosar} and ARINC653~\cite{arinc}, to isolate
tasks of different criticality levels. However, these types of systems are not
able to take advantage of legacy software, including libraries and device
drivers written for the newest hardware. In contrast, systems such as Linux,
Windows and OS X are regularly updated with features that would take an
operating system developer years to reproduce in a clean-slate
design. Unfortunately, general purpose systems lack the necessary temporal and
spatial requirements, including the ability to perform real-time sensing,
processing and actuation required by emerging mixed-criticality systems.

%%%%%%%%%%%
%% Increasingly complex embedded systems are being used in a range of
%% applications, including autonomous vehicles, advanced driver assistance
%% systems (ADAS), robotics and manufacturing. These systems often require
%% real-time task pipelines to perform sensor data processing and device I/O.
%% Additionally, they comprise hardware and software components with different
%% criticality levels. An adaptive cruise control, for example, adjusts a car's
%% speed in real-time based on the distance to other vehicles detected by radar
%% sensors.

%% While many real-time systems
%% exist~\cite{vxworks,ecos,rtems,freertos,lynx,autosar,Leiner:SAFECOMP2007} they
%% often lack the richness of functionality found in general purpose systems such
%% as Linux, Windows and OS X.
%%%%

In this paper, we present a system called Boomerang. Boomerang uses a
partitioning hypervisor \cite{jailhouse}, which separates the hardware of a
physical machine into different guest domains that {\em directly} manage their
assigned resources. This contrasts with a conventional {\em multiplexing} (or
consolidating) hypervisor, which intervenes in the sharing of physical machine
resources among multiple guests. Boomerang's approach removes the hypervisor
from resource management, once CPU cores, physical memory and I/O devices are
assigned to separate guests.

Using separate partitions, Boomerang supports the co-existence of a
real-time operating system (RTOS) and a legacy system such as
Linux. Rather than treating these systems as separate guests,
Boomerang establishes a tightly-coupled {\em symbiotic relationship},
such that the RTOS is empowered with legacy features, and the legacy
system is empowered with real-time capabilities. For example, a
Boomerang Linux partition might support OpenGL and CUDA libraries for
hardware accelerators, camera devices, and machine learning
algorithms, which would be difficult to write and certify for an RTOS.
Likewise, the RTOS partition in Boomerang provides the timing
guarantees for real-time tasks to perform sensor data processing and
actuation.

Key to this paper's contributions is the construction of a {\em
  composable tuned pipe} abstraction. This abstraction implements
  real-time task pipelines that ensure end-to-end guarantees on
  sensing, processing and actuation, spanning both RTOS and legacy OS
  services. Boomerang extends prior work on tuned pipes between a USB
  device and a task running in the same
  OS~\cite{Golchin_Cheng_West_2018} to encompass task pipelines
  spanning an RTOS and another guest. The aim is to show that
  Boomerang is able to combine legacy and real-time services in a way
  that ensures information flow is bounded by throughput, loss and
  delay constraints.
  
As stated above, many emerging mixed-criticality systems require tasks
  to process sensory inputs before subsequently generating outputs
  that affect the actuation of a device. For example, a cruise control
  system in an electric car may collect data from cameras and speed
  sensors before determining that the motors need to change speed to
  keep a safe distance to the vehicle ahead.

Novel to Boomerang's composable tuned pipes is the ability for an
integrated RTOS based on Quest~\cite{quest} to manage I/O that
requires services in a legacy system such as Linux.  We show how to
construct composable task pipelines in Boomerang that combine tasks
spanning Quest and a legacy Linux system. By assigning time-critical
I/O to Quest, Boomerang ensures that complementary services provided
by Linux meet end-to-end timing guarantees. We compare Boomerang to a
standalone Linux system, using specific cores to handle
timing-sensitive I/O. Boomerang not only benefits from spatial
isolation, it also outperforms a standalone Linux system using
deadline-based CPU reservations for pipeline tasks.  We also show how
Boomerang outperforms a partitioning hypervisor called ACRN, designed
for automotive systems.

The following section provides background to the problem addressed by
Boomerang. Section~\ref{sect:system} describes the Boomerang partitioning
hypervisor and composable tuned pipes. An evaluation of Boomerang is described
in Section~\ref{sect:eval}. Related work is discussed in
Section~\ref{sect:related}. Finally, conclusions and future work are described
in Section~\ref{sect:conc}.

\section{Background}
\label{sect:background}
Boomerang supports composable task pipelines that form a round-trip path,
originating from a device input and ultimately finishing with a device
output. It is designed specifically for applications that require sensing,
processing and actuation.

\begin{figure}[!htb]
%\vspace{-0.1in}
  \centering
  \includegraphics[width=\columnwidth]{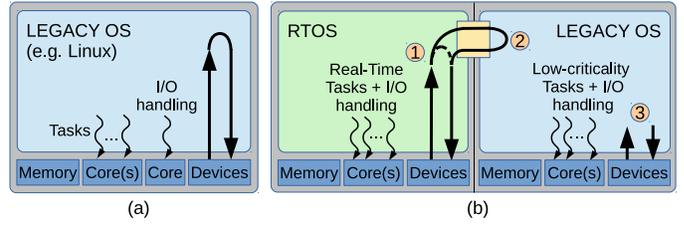}
%\vspace{-0.1in}
	\caption{(a) Round-trip I/O in a single OS, and (b) possible I/O paths
          in a Boomerang partitioning hypervisor.}
  \label{fig:boomerang_overview}
%\vspace{-0.15in}
\end{figure}

Figure~\ref{fig:boomerang_overview}(a) shows the round-trip path in a typical
OS. A device acknowledges the completion of an I/O request by generating an
interrupt. Most systems handle interrupts at priorities above those of
software tasks. They also incorrectly charge interrupt handling to the task
that was preempted by the arrival of the interrupt. Worse still, a burst of
interrupts within a short time may delay a time-critical task enough to miss
its deadline~\cite{process-aware,quest}.

Figure~\ref{fig:boomerang_overview}(a) uses a dedicated core for I/O
handling of device interrupts, to avoid interference with task
execution. However, the single OS approach does not provide adequate
spatial isolation of tasks of different criticalities, and
underutilizes the core exclusively used for interrupt handling. If the
OS malfunctions then tasks of all criticalities are potentially
compromised. In contrast, Figure~\ref{fig:boomerang_overview}(b) shows
how Boomerang supports three different classes of I/O using a
partitioning hypervisor~\cite{quest-v-toc,ramsauer2017look} to
separate highly critical timing sensitive operations from less
critical system components using different guest OSes.

In the first case (shown with a dashed line), all I/O is contained
within the RTOS. Real-time tasks and interrupt handlers for device I/O
share the same processor cores, as the RTOS ensures predictable timing
guarantees on task and I/O processing.

In the second case, the I/O path traverses a task pipeline that enters
into a legacy OS via secure shared memory. Here, the legacy OS
provides services that would require significant effort to port to the
RTOS. The round-trip I/O path in case 2 is still able to meet
end-to-end timing guarantees because the tasks in the legacy OS are
isolated from timing unpredictability caused by interrupts. This is
possible by demoting interrupts (in the legacy OS) to priorities that
are distinctly lower than those of tasks. Additionally, legacy OSes
such as Linux support {\tt SCHED\_DEADLINE} execution for tasks,
thereby ensuring some degree of timing guarantees, as long as there is
no interference from interrupts~\cite{Golchin_Cheng_West_2018}.

In the third case, it may be necessary for some I/O to be handled by a legacy
system, which has drivers and libraries that are unavailable in the RTOS. For
example, a series of cameras used in a driverless car need suitable device
drivers and machine learning algorithms to perform object classification. The
outcomes of object classification dictate whether information needs to be
communicated to the RTOS to issue real-time outputs that adjust vehicle
motion. As with the single OS approach, I/O originating in case 3 may handle
interrupts on a dedicated core, to avoid interference with tasks that
serve RTOS requests in case 2. Alternatively, I/O processing in the legacy OS
is given lower priority than task execution, leaving critical I/O to the
real-time OS.

\subsection{VCPU Scheduling}
\label{sect:scheduling}
Boomerang's partitioning hypervisor allows each guest to directly
manage its assignment of physical CPUs (PCPUs)~\footnote{A PCPU is
  either a processor core, hardware thread, or individual CPU.}. This
differs from a traditional hypervisor, which schedules guest threads
on virtual CPUs (VCPUs) and then maps those onto PCPUs. Instead,
Boomerang's guest RTOS implements its own VCPU scheduler using the
PCPUs available to it, without any need for additional scheduling in
the hypervisor. At the same time, other guests running on Boomerang
schedule threads directly on their available PCPUs.

Boomerang uses the Quest RTOS~\cite{quest} to assign a processor
capacity reserve~\cite {Mercer94processorcapacity} to each VCPU.  Each
reserve has a budget capacity, $C$, and period, $T$. A VCPU is
required to receive at least C units of execution time every T time
units when it is runnable, as long as a schedulability
test~\cite{Lehoczky:89} is passed when creating new VCPUs. This way,
the Quest scheduler guarantees temporal isolation between threads
associated with different VCPUs.

\begin{figure}[!htb]
\vspace{-0.1in}
  \centering
  \includegraphics[width=0.4\textwidth]{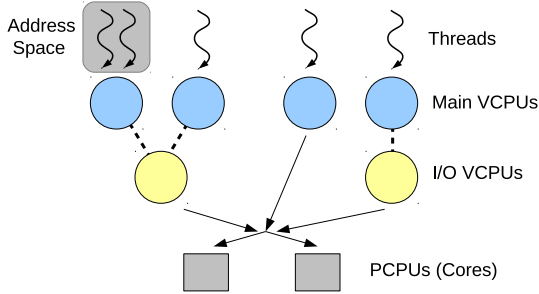}
%\vspace{-0.1in}
  \caption{VCPU scheduling hierarchy in Quest.}
  \label{fig:vcpu}
%\vspace{-0.15in}
\end{figure}

Figure~\ref{fig:vcpu} shows the scheduling of threads and VCPUs for
real-time tasks and interrupt handlers. Tasks are assigned to {\em
  Main VCPUs}, and separate {\em IO VCPUs} are used for interrupt
handling. Main VCPUs are implemented as Sporadic
Servers~\cite{spruntsporadic}. Each Sporadic Server keeps track of its
VCPU's budget usage, and constructs a list of timestamped future
replenishments, to ensure timing guarantees. By default each Sporadic
Server VCPU is scheduled using Rate-Monotonic Scheduling
(RMS)~\cite{RMS}, although an alternative policy such as
earliest-deadline first (EDF) may be chosen. With RMS, the VCPU with
the smallest period, $T$, has the highest priority.

To ensure that tasks are isolated from interrupts, Quest promotes
interrupt handling to a schedulable thread context, whose execution is
charged to an IO VCPU. Each IO VCPU is associated with the Main VCPU
that led to the occurrence of the interrupt. Such occurrences result
from tasks issuing blocking requests (e.g., via a {\tt read()} system
call), or a system thread awaiting a kernel event.

Consider a task that issues a blocking I/O request on a device (e.g.,
USB interface).  When the task blocks, it stops charging execution
time to its Main VCPU. Some time later an interrupt occurs when an I/O
transfer is complete. This causes a {\em top half} handler to execute,
which determines the Main VCPU waiting on I/O. The top half then
inserts into a system ready queue an {\em IO VCPU} with a dynamically
calculated budget and period, based on the parameters of its
corresponding Main VCPU. Finally, the interrupt is acknowledged, and
all subsequent handling occurs in a {\em bottom half} thread context,
when the corresponding IO VCPU is scheduled.  Consequently, all bottom
half execution time is charged to its IO VCPU before the blocked task
resumes execution on its Main VCPU.

Each IO VCPU in Quest is given a utilization bound,
$U_{IO}$. There is one IO VCPU for each device class, with classes
existing for USB, networking, ATA, and GPIO devices, among
others. When an IO VCPU is added to the scheduler ready queue, its
budget is set to $U_{IO}{\times}T_{Main}$ and its period is set to
$T_{Main}$, where $T_{Main}$ is the period of the Main VCPU of the
source entity associated with the interrupt. Quest is then able to
correctly schedule bottom half interrupt handlers at the priority of
the source task running on a Main VCPU. This contrasts with systems
such as Linux, which schedule bottom halves (a.k.a., tasklets or
softirqs) at priorities that are not tied to the source of
corresponding interrupts.

IO VCPUs have a dynamically calculated budget and period based on the
Main VCPUs they serve, to avoid the overhead of maintaining
replenishment lists for short-lived interrupt service routines (ISRs).
This budget is eligible for use as long as the sustained IO VCPU's
utilization does not exceed $U_{IO}$. This policy is shown to be
effective for short-lived interrupt service routines (ISRs), which
would fragment a Sporadic Server budget as used for Main VCPUs.

Quest requires reprogramming of hardware timers in {\em one-shot}
mode, to determine the next system event. This is similar to Linux's
tickless operation. As IO VCPUs only have {\em one budget
replenishment} to consider, rather than a list, this leads to reduced
timer reprogramming overhead.

\subsection{Communication Model}
\label{sect:async_comm}

Data flow involves a pipeline of communicating tasks. Each task
processes its input data to produce output, either for devices or
subsequent tasks in the pipeline. This leads to a communication model
characterized by: (1) the interarrival times of tasks in the pipeline,
(2) inter-task buffering, and (3) each tasks' access pattern to
communication buffers.

\textbf{Task Interarrival Times.} Each task ordinarily samples input data
periodically. However, a task will block if data is unavailable,
leading to aperiodic or irregular intervals between successive task
instances.  Either way, a task pipeline's timing requirements assume
that data will propagate with a minimum inter-arrival time between
tasks.

\textbf{Register-based versus FIFO-based Communication.} A FIFO-based
shared buffer is used in scenarios where {\em data history} is an
important factor. However, in sensor-data processing the most recent
data is often more important. For example, a driving assistance system
should always compute outputs that affect vehicle dynamics from the
latest sensor data. FIFO-based communication results in loosely
synchronous communication: the producer is suspended when the FIFO
buffer is full and the consumer is suspended when the buffer is
empty. Register-based communication achieves asynchrony between two
parties using Simpson's four-slot algorithm~\cite{4slots}.

\textbf{Implicit versus Explicit Communication.} Explicit communication allows
access to shared data at any time during a task's
execution. This might lead to data inconsistency in the presence of
task preemption.   Conversely, the implicit communication
model~\cite{ecrts2017comm} essentially follows a read-before-execute
paradigm to avoid data inconsistency. It mandates a task to make a
local copy of the shared data at the beginning of its execution and to
work on that copy throughout its execution.

Boomerang supports both periodic and aperiodic tasks. It also supports
both register- and FIFO-based communication. Implicit communication is
enforced for data consistency.

\section{Boomerang}
\label{sect:system}
The Boomerang partitioning hypervisor divides processor cores,
physical memory and I/O devices among guest domains. Each guest
manages its physical resources {\em without} involvement of the
hypervisor. This has two important properties: (1) the hypervisor is
only used to bootstrap the system and to establish secure
communication channels between guests using hardware extended page
tables (EPTs)~\footnote{Intel processors with VT-x capabilities refer
to these tables as EPTs. AMD-V processors have similar nested page
tables (NPTs).}, and (2) the hypervisor is removed from runtime
management of physical machine resources, making its trusted code base
extremely small.

Boomerang's partitioning hypervisor has a text segment of less than 4KB,
although more space is needed for EPTs (e.g., 24KB for a 4GB guest). Given the
hypervisor is not accessed under normal guest operation, the system's most
privileged {\em ring of protection} is less susceptible to security attacks
than a conventional OS image running directly on hardware. In the latter case,
system calls must pass control to the OS kernel, whereas in Boomerang these
are restricted to the local guest.

Unlike traditional hypervisors that {\em multiplex} guests onto the same
shared physical machine, partitioning hypervisors offer opportunities for
applications that require security and timing predictability. Hardware
virtualization features isolate guests, using an additional ring of protection
reserved for the hypervisor. At the same time, time-critical guests are able
to run real-time resource management policies without being compromised by
additional resource management policies in the hypervisor.

We see partitioning hypervisors as being suitable for mixed-criticality
systems, requiring spatial and temporal isolation of application tasks and
software components according to different system criticality levels. For
example, automotive systems adhering to standards such as ISO
26262~\cite{ISO26262-3} are required to meet specific functional safety
requirements, according to several classes of {\em automotive safety integrity
  levels} known as ASIL A-D. Software certified to ASIL D standard operates at
the most stringent safety level, where the risk of failure is potentially life
threatening. In contrast, ASIL A applies to software that has a very low
probability of significant human injury even during failures. Other standards
such as ARINC 653 and DO-178B have similar requirements for avionics
systems. For these types of systems, it is possible to assign software to
machine partitions according to their safety integrity levels.

\subsection{Composable Tuned Pipes}
\label{sect:composable_tuned_pipes}
%% Include details of composable pipes here.

\begin{figure}[!ht]
%\vspace{-0.1in}
  \centering
  \includegraphics[width=0.35\textwidth]{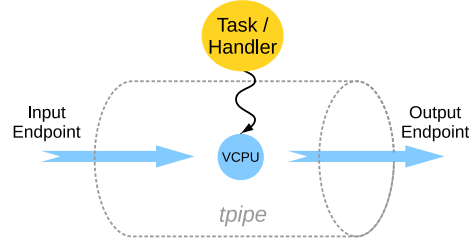}
%\vspace{-0.05in}
	\caption{A tuned pipe.}
  \label{fig:tpipe}
%\vspace{-0.1in}
\end{figure}

Figure~\ref{fig:tpipe} shows a logical representation of a single {\em tuned
pipe} (a.k.a., {\em tpipe}). A pipe has one {\em pipe processor} and two {\em
endpoints}, with one endpoint for input and the other for output.  A pipe
processor is represented by a VCPU, guaranteeing at least $C$ units of
execution time every $T$ time units when runnable. Pipe processors are
associated with tasks bound to Main VCPUs, or threaded interrupt handlers
bound to I/O VCPUs.

A tuned pipe guarantees data flowing from an input to an output
endpoint is processed according to specific service
requirements. These requirements apply end-to-end, through a {\em
pipeline} of one or more tuned pipes. If the pipeline is lossless, it
ensures specific throughput and delay guarantees, whereas if it is
lossy, it guarantees a maximum fraction of lost data while meeting
delay bounds.

Boomerang maintains a local repository for each guest OS (a.k.a., sandbox or
machine partition), which stores information about available endpoints. The
repository records a globally unique identifier for each endpoint, in the
form: {\em hostID:sandboxID:asID:epID}. This distinguishes endpoints in
different host machines (by {\em hostID})~\footnote{In this paper, we restrict
communication within the same host machine.}, sandboxes (by {\em sandboxID}),
and address spaces (by {\em asID}). Access capabilities restrict which tuned
pipes are able to connect to endpoints.

The rules controlling connectivity to endpoints are a topic of ongoing
research. They have implications for secure information flow
analysis~\cite{zeldovich06histar,efstathopoulos05asbestos,bpl}, which
is outside the scope of this paper. Notwithstanding, pipelines may be
constructed within a single address space, between address spaces in
the same machine partition, between different partitions on the same
host, and across different hosts.

When creating a tuned pipe, Boomerang automatically calculates (i.e., {\em
tunes}) the budget and period of the pipe VCPU to ensure end-to-end guarantees
are met. Tuned pipes are created with a call to {\em tpipe()}, as follows:

\footnotesize
\begin{verbatim}
tpipe_id_t tpipe(ep_t *inp[], int n_inp, ep_t *outp,
           qos_t spec, tpipe_task_t func, void* arg);
\end{verbatim}
\normalsize

The input endpoint of the new tuned pipe specifies an array of
pointers, {\tt inp}, to endpoint types. This array identifies the
endpoint addresses of {\tt n\_inp} inputs to the tuned pipe, along
with the buffering semantics of each input, which will be discussed in
Section~\ref{sect:POSIX-comparison}.

Data flowing into the tuned pipe is processed by a specific callback
function ({\tt func}), which sends its output to specific destinations
connected to the output endpoint, identified by {\tt outp}. The
callback function takes an optional argument ({\tt arg}), and runs in
its own thread context. The thread context defines a task, which is
bound to a VCPU having an automatically-generated budget, $C_i$, and
period, $T_i$, for the tuned pipe, $tpipe_i$. The budget and period
are derived from the quality-of-service (QoS) requirement ({\tt
spec}) for end-to-end throughput and delay on data processing.  This
requirement must also satisfy the schedulability of all VCPUs on a
given physical CPU (PCPU), otherwise the tuned pipe is not created. If
a tuned pipe is successfully created, it is given a unique ID within
its guest OS.

$tpipe_i$ requires its callback function to process data from one or
more input endpoints and produce output in one quantum of size $C_i$,
every period, $T_i$. Functions are selected from a predefined
repository of callbacks. Each callback has a known worst-case
execution time (WCET) based on pre-profiled timing information to
handle a maximum $I_i$ inputs and produce up to $O_i$ outputs in one
quantum. The actual amount of processing in a quantum depends on the
availability of data in input buffers, and how many outputs need to be
written.

Each function in the repository declares the allowable buffering
capabilities for its inputs and outputs. Any tuned pipe connecting to
another with a function that does not match the allowed buffering
capabilities is rejected.

\subsection{POSIX Pipes versus Tuned Pipes}
\label{sect:POSIX-comparison}
Similarities exist between a pair of tuned pipes and a single POSIX
pipe. The latter provides a shared memory buffer that is accessible to
a group of communicating threads via file descriptors. The file
descriptors describe the endpoint capabilities, including whether the
pipe is readable or writable.

\begin{figure}[!ht]
%\vspace{-0.15in}
  \centering
  \includegraphics[width=0.47\textwidth]{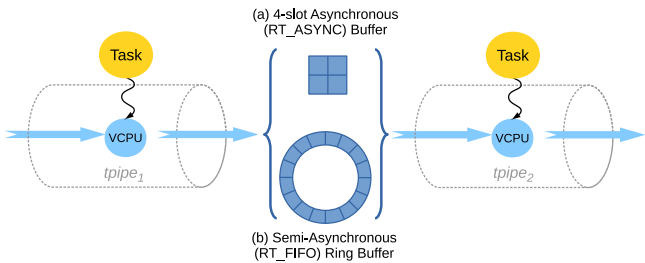}
%\vspace{-0.15in}
	\caption{A two-stage pipeline with (a) 4-slot asynchronous buffering,
  and (b) semi-asynchronous ring buffering.}
  \label{fig:endpoint_buffering}
%\vspace{-0.1in}
\end{figure}

A tuned pipe pair in Boomerang differ from a POSIX pipe by capturing
the timing requirements for data processing and communication. They
also define the buffering semantics for I/O endpoints. Two pipes,
$tpipe_i$ and $tpipe_j$ are composed by connecting the output endpoint
of $tpipe_i$ to the input endpoint of $tpipe_j$. Boomerang allows the
composition of two or more pipes to support either asynchronous ({\tt
RT\_ASYNC}) or semi-asynchronous ({\tt RT\_FIFO}) communication, as
shown in Figure~\ref{fig:endpoint_buffering}.

With {\tt RT\_ASYNC}, Simpson's Four Slot buffering
scheme~\cite{4slots,rushby2002model} is used to allow the two pipe
threads to execute independently of each other. Four Slot
communication guarantees {\em freshness} and {\em integrity} of data
objects exchanged between a producer and consumer, without the sender
or receiver ever having to block. Freshness guarantees the most recent
value of a data object is made available. Integrity ensures a data
object is not partially updated before the previous object has been
read in entirety.

With {\tt RT\_FIFO}, a ring-buffer is established between the
communicating pair of pipes to avoid data loss.  However, the sender
must block when the buffer is full, and the receiver must block when
the buffer is empty. This places a timing dependency on producers and
consumers, which potentially violates end-to-end timing guarantees
unless data flow rates are managed correctly.

\subsection{Device versus Task Pipes}
\label{sect:pipe-types}
Boomerang's RTOS provides a pre-defined set of tuned pipes for all
devices involved in real-time I/O. A {\em device pipe} features an IO
VCPU for interrupt handling, and an optional Main VCPU for endpoint
buffer management of shared devices. Sharing requires scatter-gather
functions to move data between the device endpoint buffer and
pipe-specific buffers of {\em task pipes}. If a device is not shared,
its handler directly accesses the buffer of a specific task pipe.

The {\tt tpipe()} call, described earlier, creates a {\em task
pipe}. Unlike a device pipe, there is no IO VCPU for interrupt
handling. Task pipes form pipelines between device pipes that act as
the sources and sinks of input and output data, respectively.

%% As with a device pipe, a task pipe is shareable with other
%% tuned pipes that have access rights. However, the data output from a
%% task pipe is {\em replicated} among all buffers that connect to
%% subsequent tuned pipes. In contrast, the data output from a device
%% pipe is sent only to the task pipe that should receive that
%% data.

\begin{figure}[!ht]
%\vspace{-0.15in}
  \centering
  \includegraphics[width=0.47\textwidth]{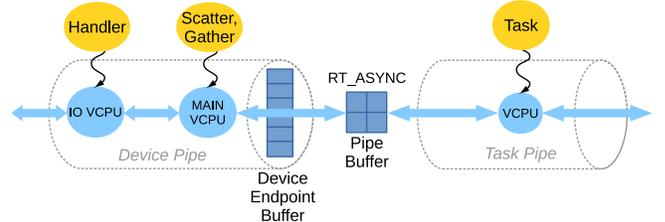}
%\vspace{-0.1in}
	\caption{Example composition of a device and task pipe for
  asynchronous I/O.}
  \label{fig:device_endpoint}
%\vspace{-0.1in}
\end{figure}

Figure~\ref{fig:device_endpoint} shows an example composition of a device and
task pipe for asynchronous (non-blocking) I/O communication. The device is
assumed to be shared with other tasks. If a task requires semi-asynchronous
device communication for blocking I/O, it would replace the four slot pipe
buffer with a ring buffer.

\subsection{Pipeline Construction}
\label{sect:pipeline_construction}

Pipelines of tuned pipes are constructed in the order in which data
flows, from input to output. A tuned pipe is responsible for the
creation of all buffers that connect to its input endpoint. It also
declares its output endpoint, which includes a count of the number of
outputs it handles. A pipeline is incomplete until all $I_i$ inputs
and $O_i$ outputs of each $tpipe_i$ are connected.

The output endpoint of each task pipe has a connection to a default device
pipe, which could be a null device. A system call interface allows this output
endpoint to be redirected to one or more different device pipes.

Once fully connected, the system activates the pipeline by allowing
each tpipe task to be scheduled for execution. Those tasks that
execute in the RTOS are runnable when they have available budgets on
their corresponding VCPUs. Tuned pipe tasks that execute in Linux are
runnable when they have available budgets in their {\tt
SCHED\_DEADLINE} scheduling class. Linux's {\tt SCHED\_DEADLINE}
scheduling class uses a Constant Bandwidth Server~\cite{cbs} to limit
the maximum CPU bandwidth consumed by a task within a specific
period. The end of the period is used to define the task deadline, and
all tasks are scheduled earliest deadline first. However, interrupt
handlers are not managed in this scheme.

Boomerang runs our in-house RTOS in one sandbox, and Linux in another
sandbox on the same physical machine. A Linux kernel module maps a
secure shared memory region by calling into the hypervisor. The
hypervisor uses EPTs to map machine physical memory into each sandbox
so they are able to communicate.

Each sandbox is equipped with kernel services that manage a local
repository of endpoints and tuned pipes. Communication services allow
queries to a remote sandbox, to discover endpoints and to connect or
disconnect from tuned pipes. Mailbox channels are established by
Boomerang to enable OSes in different sandboxes to send remote OS
requests. Access policies determine whether address spaces in the
local or remote sandbox are able to connect to endpoints of existing
tuned pipes.

Boomerang's RTOS provides a remote shell to Linux through
inter-sandbox shared memory. Linux uses a kernel module to allow
user-space application programs with root privilege to execute shell
commands on the RTOS. A shell interface allows pipelines of tuned
pipes to be constructed. The RTOS is able to query endpoints and tuned
pipes that exist in Linux, and issue requests to connect to them via
{\tt tpipe()} calls.

After the construction of the pipeline, the RTOS runs an end-to-end
throughput and delay analysis. If the end-to-end requirements are met
for the pipeline, the transmission of data is allowed to begin from
the RTOS.  Tuned pipe functions synchronize their start and end of
operation life-cycle using Start-Of-Task and End-Of-Task packets on
their input endpoints.

The following example illustrates a pipeline specification:
\begin{equation*}
[*](A~|~B), C~|~D~|~E,~F~ [e2e\_tput~|~loss\_rate,~e2e\_delay]
\end{equation*}

\begin{figure}[!ht]
%\vspace{-0.1in}
  \centering
  \includegraphics[width=0.47\textwidth]{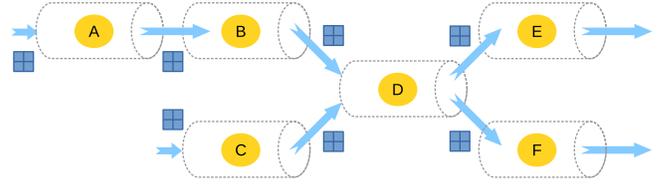}
%\vspace{-0.05in}
	\caption{An example pipeline with multiple inputs \& outputs.}
        \label{fig:pipeline_example}
%\vspace{-0.05in}
\end{figure}

The resultant pipeline is shown in Figure~\ref{fig:pipeline_example}.
Boomerang's repository of tuned pipe functions requires that $A$ and
$C$ connect to a device output endpoint for reading, while $E$ and $F$
connect to a device input endpoint for writing.

Boomerang defaults to non-blocking tuned pipe semantics, where data
freshness is more important than lossless
communication. Figure~\ref{fig:pipeline_example} shows four-slot
buffering of all pipeline stages. If lossless communication is
required, the entire pipeline specification is preceded by an
asterisk. This pipeline would then use FIFO buffers between each pair
of tuned pipes.

With four-slot buffering, the entire pipeline has an optional
end-to-end service specification in terms of tolerable $loss\_rate$
and $e2e\_delay$. With FIFO buffering, the pipeline is specified with
an optional end-to-end throughput, $e2e\_tput$, and delay. The
throughput is measured as the minimum number of {\em data objects} per
unit time exiting a final tuned pipe, while the delay is measured in
microseconds. Each data object represents a message, which is the size
of one slot of either a four-slot or FIFO buffer.

If the QoS specification is omitted, then the pipeline defaults to
best effort. In such case, the VCPUs of each tuned pipe revert to
their default values. If the pipeline overloads the PCPUs to which it
is assigned, leading to an infeasible schedule, its VCPU periods are
repeatedly extended until the pipeline is schedulable.

The shell interpreter allows parallel sections of a pipeline to be
defined by comma-separated lists of tuned pipes. Here, the pipeline
section $A~|~B$ runs parallel with $C$. This could be representative
of two separate input sensor streams coming from two different
devices. Parentheses ensure correct grouping of pipeline sequences, while
two tuned pipes are connected using the shell vertical bar symbol ($|$).

In the example, the outputs of $B$ and $C$ feed into the single tuned
pipe, $D$. Similarly, the output of $D$ is split across $E$ and
$F$. $D$ might represent a sensor fusion and control task, while $E$
and $F$ might be specific actuator tasks that output their data to
different devices. In an automotive system, for example, $E$ and $F$
might send their outputs to two different CAN buses, managed by device
pipes.

The $e2e\_delay$ constraint applies to the longest path through the
pipeline, while for FIFO-buffered communication the $e2e\_tput$
applies to whichever final task pipe has the lowest rate of output. If
FIFO-buffering were used in the figure, whichever of $E$ and $F$ had
the lowest output rate would dictate the end-to-end throughput.

As a four-slot buffered pipeline allows each tuned pipe to read and process
whatever data sits in its input buffers, it is possible that new data has
overwritten old data before the consumer runs. This happens if the producer
has an arrival rate, $\lambda=1/T_{p}$, greater than the consumer's service
rate, $\mu=1/T_{c}$. Here, it is assumed that $T_{p}$ and $T_{c}$ are set to
ensure one message transfer every corresponding period, regardless of whether
it is a new message or not.

%% We assume that *one* message is propagated by each pipeline stage
%% every period of the corresponding tuned pipe. The analysis works fine
%% when we're dealing with 4-slot buffers, because we consume whatever is
%% in the buffer, even if it is not new (in which case we reuse what we
%% saw before). However, with FIFO we will block if the producer and
%% consumer rates are not matched, as the buffer will either fill up (if
%% T_consumer > T_producer) or empty (if T_consumer < T_producer).

%% A FIFO consumer has to remove data from its buffer to make space for
%% new data. With 4-slot, data is either over-written by a fast producer,
%% or re-read more than once by a fast consumer.

%% We cannot guarantee that a pipeline with FIFO buffers can process a
%% message every period of the corresponding tuned pipe. I think the
%% e2e-delay becomes [sum(i=1 to n) T_i x ceiling(max(T)/T_i)] where n is
%% the number of tuned pipe stages and max(T) is the tuned pipe with the
%% largest period. E2e-throughput becomes: 1/max(for all i)[Ti x
%% ceiling(max(T)/T_i]. The worst-case period has to round up the periods
%% of other tuned pipes that might be phase-shifted.

%% Things get more complicated with a mix of FIFO and 4-slot buffers, so
%% I need to think some more.

\subsection{End-to-end QoS Guarantees}
\label{sect:qos}
Given a pipeline of tuned pipes and buffers, Boomerang runs a
constraint solver to determine $C_i$ and $T_i$ for each $tpipe_i$. The
function executed by $tpipe_i$ is assumed to process at least one of
its $I_i$ inputs and generate one of its $O_i$ outputs every period,
$T_i$. Essentially, one or more processed data {\em messages}
propagate through a tuned pipe within $C_i$ execution time. Boomerang
assumes that $C_i$ is derived by pre-profiling the WCET of the
corresponding task function. This WCET is then stored in the local
repository, along with the set of inputs and outputs used by the
function.

For a pipeline to successfully meet its end-to-end timing
requirements, Boomerang must still determine each period,
$T_i~|~T_i{>}C_i$, and possibly scale each service time $C_i$ to
forward more than one message at a time. It follows that a
FIFO buffered pipeline successfully meets its end-to-end timing
requirements if:

\begin{enumerate}
\item $\sum_{i{\in}l}T_i{\leq}e2e\_delay$, for the longest path $l$,
\item $min_{\forall{i}}\{\frac{m_i}{T_i}\}{\geq}e2e\_tput$, where
$m_i{\geq}1$ messages are\\ transferred by $tpipe_i$ every $C_i$,
\item all FIFO buffers are sized to ensure no additional
blocking delays of tasks, and
\item all task scheduling constraints are satisfied on their respective PCPUs.
\end{enumerate}

Similarly, a pipeline with four-slot buffering meets its end-to-end
requirements if:

\begin{enumerate}
\item $\sum_{i{\in}l}T_i{\leq}e2e\_delay$, for the longest path $l$,
\item $max\{1-\frac{T_{p}}{T_{c}}\}{\leq}{loss\_rate}$, for all
$T_{p}{\leq}T_{c}$, and
\item all task scheduling constraints are satisfied on their respective PCPUs.
\end{enumerate}
The end-to-end delay represents the time for a message to traverse the
longest path through a pipeline. The final message output from the
pipeline is a transformation of data propagated through each tuned
pipe.

The worst-case end-to-end delay is the sum of all the periods of the
tuned pipes in the longest path, plus any blocking delays. The
blocking delays are zero with asynchronous communication as each tuned
pipe processes its most recent data, regardless of it being
updated. Similarly, blocking delays are avoided with FIFO-based
communication if each buffer is never empty or totally full.

It follows that each tuned pipe propagates a message after $C_i$
worst-case execution time. However, if data arrives at the inputs to a
tuned pipe when it has just depleted its budget, it must wait
$T_i-C_i$ before the budget is replenished. If the next tuned pipe
is not synchronized to start exactly when the previous pipe
forwards its data there could be an additional delay of $T_i-C_i$ on
top of $C_i$ to process the data in $tpipe_i$.

\begin{figure}[!ht]
%\vspace{-0.15in}
  \centering
  \includegraphics[width=0.35\textwidth]{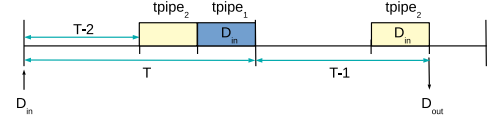}
%\vspace{-0.1in}
	\caption{Worst-case delay:
	$D_{in}{\shortrightarrow}tpipe_1{\shortrightarrow}tpipe_2{\shortrightarrow}D_{out}$.}
  \label{fig:example_schedule}
%\vspace{-0.1in}
\end{figure}

To see this more clearly, consider a system of $T$ tasks each with a
service time of $1$ time unit every $T$. Suppose two of these tasks
are associated with $tpipe_1$ and $tpipe_2$. Input data $D_{in}$ to
$tpipe_1$ is processed and forwarded to $tpipe_2$, which produces
$D_{out}$. These two tuned pipes form a pipeline, while all other
tasks compete for execution on the same PCPU. Using either rate
monotonic or earliest deadline first scheduling~\cite{RMS} yields the
same schedule in this case: neglecting scheduling overheads, each task
has the same priority. A possible schedule is shown in
Figure~\ref{fig:example_schedule}.

The worst-case end-to-end delay is when each of the $T-2$ tasks other than
those for $tpipe_1$ and $tpipe_2$ run immediately after the data, $D_{in}$,
has arrived. Then, $tpipe_{2}$ executes and processes old input data before
$tpipe_{1}$ is able to read $D_{in}$. Consequently, $tpipe_{1}$ does not
process $D_{in}$ and forward the output to $tpipe_{2}$ until $T$ time after
the data first arrived. Similarly, $tpipe_{2}$ is not able to run again until
$2T-2$, when it finally reads $D_{in}$. This is because the scheduler will not
provide it with a budget replenishment until one period after it last
executed. The total end-to-end delay between $D_{out}$ and $D_{in}$ is
therefore $2T-1$. For large $T$ this approaches a worst-case delay of
$2T$. Extending this to more than two tasks in a pipeline leads to the
worst-case end-to-end delay being the sum of the corresponding tuned pipe
periods.

%%%%%%

The end-to-end throughput of a path through a FIFO buffered pipeline
is limited by the minimum output rate of any one tuned pipe in that
path. A tuned pipe's output rate is how many messages it is able to
forward in its period. As FIFO buffering allows $tpipe_i$ to forward
$m_i{\geq}1$ messages per period, the minimum value of
$\frac{m_i}{T_i}{\geq}e2e\_tput$ for all $i$ is a
lower-bound on overall throughput.

For any pair of tuned pipes connected via FIFO buffers, it is essential that
blocking delays are factored into the end-to-end service guarantees. Boomerang
tries to avoid blocking on message exchanges by matching the arrival and
departure rates of messages passed through shared FIFO buffers.

Suppose a producing and consuming pair of tuned pipes have budgets
$C_{p}$ and $C_{c}$, respectively. Given $C_{p}=C_{in}$ is sufficient
to produce one message in $T_{p}$, and $C_{c}=C_{out}$ is sufficient
to consume one message in $T_{c}$, Boomerang starts by setting
$T_{p}=T_{c}=\Delta$, where $\Delta{\cdot}n=e2e\_delay$, and $n$ is
the number of tuned pipes in the longest path. This ensures the
producer and consumer are {\em rate-matched}, to prevent the
buffer between them either completely filling or emptying.

Rate-matching is applied to all tuned pipes in the pipeline. If
the pipeline cannot feasibly be scheduled on its PCPUs, each tuned
pipe period is scaled by a factor $\alpha$, where
$\alpha > 1$. This is repeated until all tuned pipes are
schedulable, but leads to a violation of the end-to-end latency
requirement.

To reduce end-to-end latency, Boomerang adjusts tuned pipe periods,
starting with the inputs to the pipeline. For each tuned pipe pair,
$T_{p}$ is repeatedly halved and $C_{c}$ is similarly doubled,
ensuring that $T_{p}{>}C_{p}$, $T_{c}{>}C_{c}$ and all VCPUs are
schedulable when possible. The doubling of $C_{c}$ enables it to
process multiple messages, $m_c$, in one budget cycle. $T_{p}$ is
reduced until either the entire pipeline meets its end-to-end delay
requirement or it is set as low as feasibly possible. If
$\sum_{i{\in}l}T_i{\leq}e2e\_delay$ for longest path $l$, the
algorithms stops, or else it moves onto the next stage in the path,
and repeats the above procedure.

If all stages of the pipeline have been processed from input to
output, the algorithm revisits each consumer whose budget is set to
process multiple messages in one period.  For each consumer, both
$C_{c}$ and $T_{c}$ are halved, as long as $C_{c}$ is no smaller than
the time to process one message. If the path's $e2e\_delay$ is
satisfied, or tuned pipe periods and budgets cannot be reduced
further, the algorithm stops. At this point each
$C_{p}=m_p{\cdot}C_{in}$ and each $C_{c}=m_c{\cdot}C_{out}$, for
$m_p,m_c{\geq}1$.

If a feasible pipeline schedule is found, each FIFO buffer is set to
have enough space for
$m_{p}{\cdot}(\lceil\frac{T_{c}}{T_{p}}\rceil+1)$ messages from the
producer. $\lceil\frac{T_{c}}{T_{p}}\rceil$ accounts for the maximum
number of times the producer can generate $m_p$ messages within one
period of the consumer. An additional $m_p$ messages may be generated
by the producer by the time the consumer accesses the buffer, due to
potential phase shifting between the two tasks.

For four-slot communication, if the consumer has a smaller period than a
producer at any stage in the pipeline, then the consumer will always see the
most recent data. Given that four-slot communication restricts each tuned pipe
to read, process and write one message every period, it is impossible for a
pipeline to lose any data if all consumer periods are smaller than their
corresponding producer periods. However, if a consumer has a larger period
than its producer, such that $T_c>T_p$, then the producer may overwrite data
before the consumer sees the previous message. It follows that the loss-rate
through a four-slot pipeline is limited to the maximum value of
$1-\frac{T_p}{T_c}$ of any stage in the pipeline. This is an important metric
for sensor data processing, where the fraction of lost data must be
constrained.

Irrespective of four-slot or FIFO-based communication, all VCPUs serving all
tuned pipes in a pipeline must satisfy the system scheduling requirements. For
$n$ tuned pipes scheduled using rate-monotonic scheduling, the scheduling
constraint is satisfied if
$\sum_{i=1}^{n}\frac{C_i}{T_i}{\leq}n{\cdot}(2^{1/n}-1)$. If earliest-deadline
first scheduling is used, the scheduling constraint is satisfied if
$\sum_{i=1}^{n}\frac{C_i}{T_i}{\leq}1$ on a single PCPU. Boomerang applies
these constraints, including utilization bounds on IO VCPUs used by device
pipes, to ensure pipeline schedulability. This holds for pipelines
encompassing our RTOS and Linux SCHED\_DEADLINE tasks.

\section{Evaluation}
\label{sect:eval}
\begin{figure}[!htb]
  \centering
  \includegraphics[width=0.8\columnwidth]{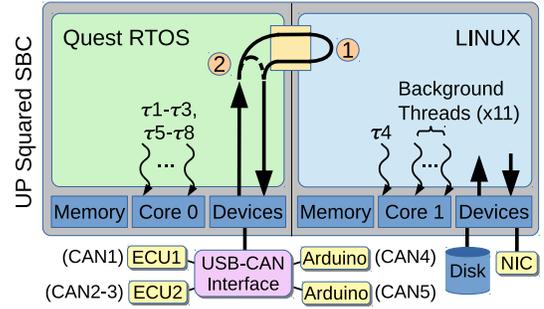}
  \caption{Boomerang experimental setup.}
  \label{fig:experimental_setup}
%  \vspace{-0.22in}
\end{figure}

We evaluated Boomerang on an Up Squared single-board computer (SBC),
featuring an Intel Pentium N4200 processor, as shown in
Figure~\ref{fig:experimental_setup}. We connected a five-channel Kvaser
USBCan Pro 5xHS CAN bus interface via USB 3.0, to emulate an
automotive system.

Traffic on CAN channels 1-3 (CAN1-3) was produced by Woodward MotoHawk
ECM5634-70 ECUs, as used in chassis and powertrain applications in a
real vehicle. Each of these channels produced data at 20\%, 30\% and
40\% of their 500kbps bandwidths, respectively.  Channels 4 and 5
(CAN4-5) were replaced with Arduino UNOs~\cite{arduinohomepg} equipped
with CAN shields, to collect performance data.

Two separate pipelines were constructed for CAN4 and CAN5, with thread
budgets and periods shown in Table~\ref{tab:pipelines_details}. These
pipelines shared three device I/O threads: {\tt mhydra\_rx} and {\tt
  mhydra\_tx} for Kvaser USBCan scatter-gather functionality, and a
USB xHCI bottom half handler ({\tt USB\_BH}). Pipeline 1 (labeled
{\circled{1}} in Figure~\ref{fig:experimental_setup}) consisted of
three task pipes: {\tt CanRead}, {\tt ProcData} \& {\tt CanWrite}, to
read, process, and write CAN data, respectively. All tasks ran in the
RTOS except {\tt ProcData ($\tau{4}$)}, which executed in Linux and
represented a task requiring capabilities unavailable in the RTOS.
Pipeline 1 extended from the RTOS into Linux via a secure shared
memory channel using extended page table mappings.  Pipeline 2 (whose
I/O path is shown with a dashed line and labeled {\circled{2}})
consisted of two task pipes that both ran in the RTOS. These tasks
were {\tt RTFusion} and {\tt RTControl}, for sensor data fusion and
control.

\footnotesize
\begin{table}[!ht]
	\centering
	\begin{tabularx}{\columnwidth}{ |p{1.0cm} p{0.35cm}|p{1.45cm}|p{1.4cm}|X|p{0.5cm}| }\hline 
		\textbf{Thread} & & \textbf{Budget} (ms) & \textbf{Period} (ms) & 
		\textbf{Utilization}~(\%) & \textbf{Core} \\\hline\hline
		\multicolumn{6}{|c|}{\textbf{Pipeline 1} (CAN4:~$\tau{1}{\shortrightarrow}\tau{2}{\shortrightarrow}\tau{3}{\shortrightarrow}\tau{4}{\shortrightarrow}\tau{5}{\shortrightarrow}\tau{6}{\shortrightarrow}\tau{1}$)} \\\hline
		\emph{USB\_BH} & ($\tau{1}$) & 0.1 & 1 & 10 & 0\\\hline
		\emph{mhydra\_rx} & ($\tau{2}$) & 0.2 & 1 & 20 & 0\\\hline
		CanRead & ($\tau{3}$) & 0.1 & 2 & 5 & 0\\\hline
		ProcData & ($\tau{4}$) & 0.2 & 2 & 10 & 1\\\hline
		CanWrite & ($\tau{5}$) & 0.1 & 2 & 5 & 0\\\hline
		\emph{mhydra\_tx} & ($\tau{6}$) & 0.2 & 1 & 20 & 0\\\hline
		\emph{USB\_BH} & ($\tau{1}$) & 0.1 & 1 & 10 & 0\\\hline\hline
		\multicolumn{6}{|c|}{\textbf{Pipeline 2} (CAN5:~$\tau{1}{\shortrightarrow}\tau{2}{\shortrightarrow}\tau{7}{\shortrightarrow}\tau{8}{\shortrightarrow}\tau{6}{\shortrightarrow}\tau{1}$)} \\\hline
		\emph{USB\_BH} & ($\tau{1}$) & 0.1 & 1 & 10 & 0\\\hline
		\emph{mhydra\_rx} & ($\tau{2}$) & 0.2 & 1 & 20 & 0\\\hline
		RTFusion & ($\tau{7}$) & 0.1 & 2 & 5 & 0\\\hline
		RTControl & ($\tau{8}$) & 0.1 & 2 & 5 & 0\\\hline
		\emph{mhydra\_tx} & ($\tau{6}$) & 0.2 & 1 & 20 & 0\\\hline
		\emph{USB\_BH} & ($\tau{1}$) & 0.1 & 1 & 10 & 0\\\hline\hline
            	\emph{Background} & $\times$11 & -- & -- & 57 & 1\\\hline
	\end{tabularx}
\caption{Pipeline task parameters.}
\label{tab:pipelines_details}
\end{table}
%\vspace{-0.1in}
\normalsize

xHCI device interrupts were mapped to Core 0, while all other device
interrupts were redirected to Core 1. 11 background tasks running on
Core 1 generated disk and network I/O activity. These included five
{\tt wget} tasks that each retrieved a copy of a 1.9GB binary image
over the Internet. Five other tasks performed file copies of a
local version of the binary image to different directories. A periodic
task additionally consumed 20\% of the CPU time to bring the total
utilization on Core 1 (including {\tt ProcData}) to 67\%.

Given the above setup, we compared Boomerang to tuned pipes
implemented in a standalone Linux SMP system. The standalone system
did not have the support of Quest, instead mapping all tasks in
Table~\ref{tab:pipelines_details} to the specified cores of the same
OS. Yocto Linux (Pyro release), featuring kernel 4.9.99 with the
PREEMPT\_RT patch, was used in both the standalone system and
Boomerang Linux guest. With Linux SMP, all threads were assigned
budgets and periods within the {\tt SCHED\_DEADLINE} scheduling class
except the {\tt USB\_BH} bottom half handler.

All experiments featuring Boomerang and Linux SMP were run for 30s,
averaged over 10 runs each. End-to-end delay results are limited to
the first 200 packet transmissions, due to space. In all results,
similar behavior was observed for more extensive runs.
	
\subsection{Asynchronous Communication}
\label{sect:async}
Asynchronous communication has the potential to suffer information
loss. We constructed two experiments with expected pipeline losses of
0\% and 20\%.  In both cases, packets for Pipelines 1 and 2 arrived
and departed on CAN4 and CAN5 channels, respectively. We measured the
round-trip time for each packet to be read from and written to each of
these channels. From Table~\ref{tab:pipelines_details} ({\tt Period}
column), the expected end-to-end delay for Pipeline 1 is 10ms, and for
Pipeline 2 is 8ms.

\subsubsection{End-to-end Delay}

\begin{figure}[!htb]
	\centering
	\begin{subfigure}{0.24\textwidth}
		\centering
		\includegraphics[width=\textwidth]{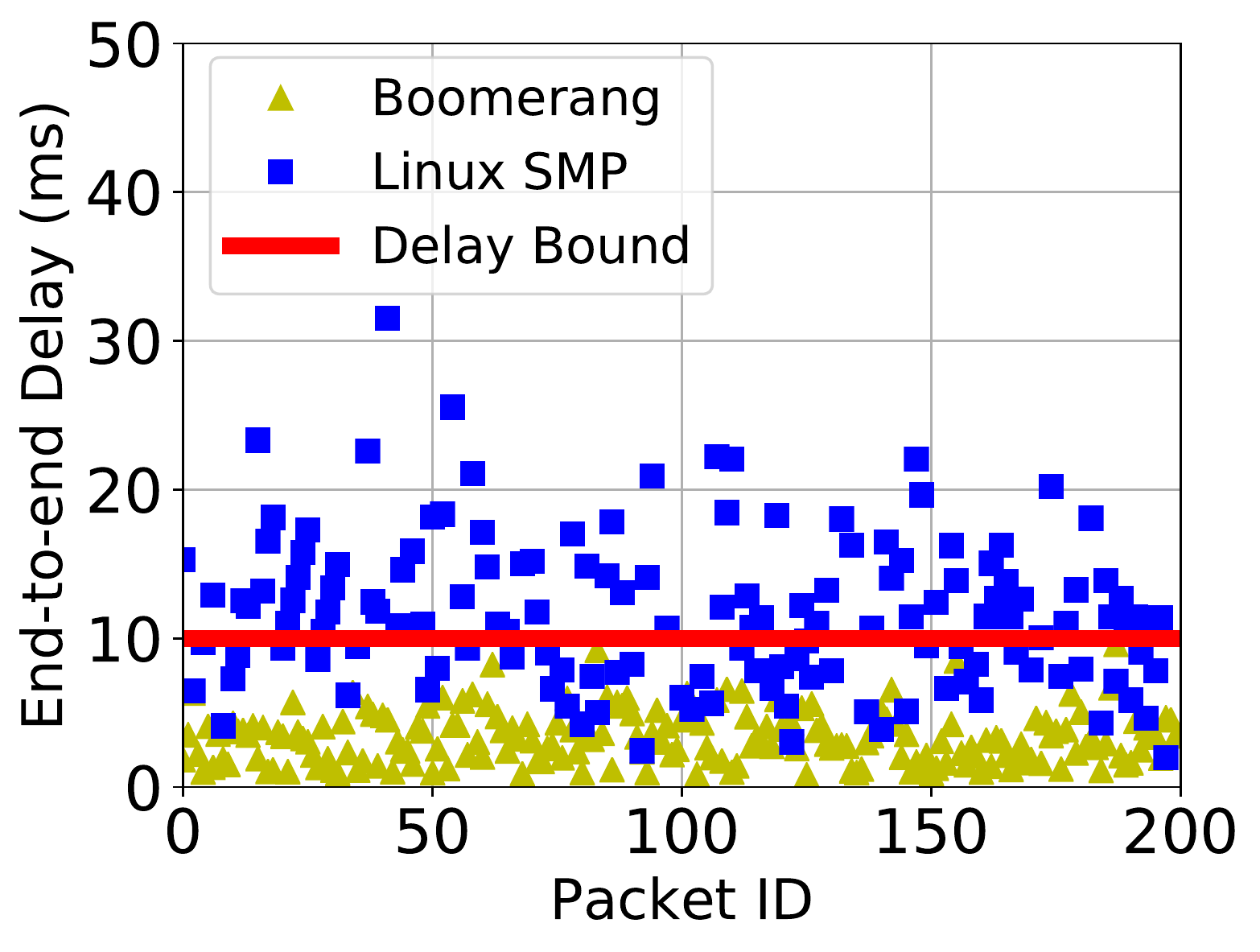}
		\caption{Pipeline 1}
		\label{fig:async_0_pipe1}
	\end{subfigure}
	\hfill
	\begin{subfigure}{0.24\textwidth}
		\centering
		\includegraphics[width=\textwidth]{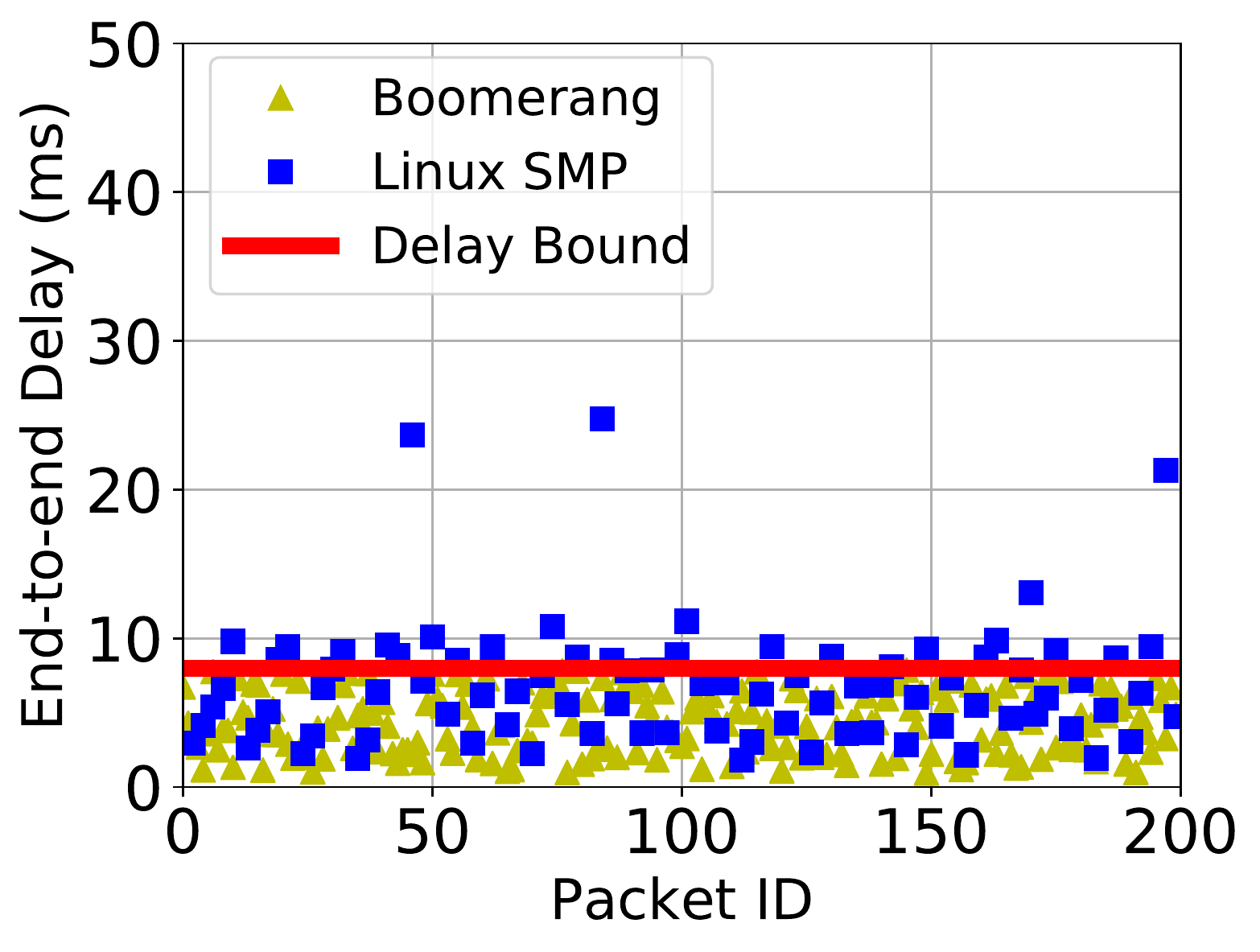}
		\caption{Pipeline 2}
		\label{fig:async_0_pipe2}
	\end{subfigure}
        \caption{End-to-end delay with no expected loss.}
%        \vspace{-0.22in}
\end{figure}

Figures~\ref{fig:async_0_pipe1} and ~\ref{fig:async_0_pipe2} show the
end-to-end delay for Pipelines 1 and 2, when there is no expected
loss. The horizontal lines represent the expected latency as
calculated from the sum of the task periods. The end-to-end latency
for Boomerang is always less than the theoretically calculated bound.
However, Linux SMP frequently fails to meet the end-to-end delay
requirements.  The main reason is the priority mismatch between
bottom-half handlers and the task awaiting I/O operations. Our RTOS
ensures that bottom-half handlers run at the correct priority with a
specific CPU reservation. Therefore, Boomerang achieves temporal
isolation between tasks and interrupts.

As Linux is unable to achieve the same level of timing guarantees,
even when tasks are guaranteed CPU reservations, there are some lost
packets as observed by the missing data points in
Figures~\ref{fig:async_0_pipe1} and
~\ref{fig:async_0_pipe2}. Table~\ref{tab:async_0} summarizes the
end-to-end latency results. It also shows that Linux suffers packet
losses of 28\% and 56\% for Pipelines 1 and 2, respectively.

\footnotesize
\begin{table}[!ht]
	\centering
		\begin{tabularx}{\columnwidth}{ |p{1.3cm}|X|X|X|X| }\hline
			\textbf{System} & \textbf{Min} (ms) & \textbf{Max} (ms) & 
			\textbf{Avg}  
			(ms) & 
			\textbf{Loss} (\%) \\\hline\hline
			\multicolumn{5}{|c|}{\textbf{Pipeline 1} (Delay bound = 10 ms)}\\\hline
			Boomerang & 0.79 & 9.57 & 3.27 &  0 \\\hline
			Linux SMP & 2.1 & 31.5 & 11.70 & 28 \\\hline\hline
			\multicolumn{5}{|c|}{\textbf{Pipeline 2} (Delay bound = 8 ms)}\\\hline
			Boomerang & 0.92 & 7.97 & 4.35 &  0 \\\hline
			Linux SMP & 1.8 & 24.77 & 6.79 & 56 \\\hline
		\end{tabularx}
		\caption{Latency - no expected loss.}
		\label{tab:async_0}
%                \vspace{-0.1in}
\end{table}
\normalsize

%% Figure~\ref{fig:async_0_int} shows the cumulative interrupts received by each
%% core with Boomerang and Linux SMP over the duration of 200 packet
%% transfers.
Boomerang experienced a total of 20623 interrupts compared to 16693
for Linux SMP during these experiments. Linux has fewer overall
interrupts but more on Core 0. We conjecture this is caused by local
APIC timer interrupts, which are influenced by the budget management
of SCHED\_DEADLINE tasks. However, this requires further
investigation. Notwithstanding, Linux SMP fails to meet end-to-end
delay guarantees because of its unpredictability in interrupt
handling.

\subsubsection{Loss}

Sensor data processing is often tolerant of lost samples. We increased the periods of certain pipeline tasks, as shown in
Table~\ref{tab:async_loss}, to allow up to 20\% lost data.
The expected latency for Pipeline 1 is now changed from 10ms to 11ms
due to increased periods of {\tt ProcData} and {\tt
CanWrite}. Similarly, the expected latency of Pipeline 2 is changed
from 8ms to 8.5ms due to the increased periodicity of {\tt RTControl}.

\footnotesize
\begin{table}[!ht]
	\centering
	\begin{tabularx}{\columnwidth}{ |X|p{2.2cm}|X|X| }\hline
		\textbf{Task} & \textbf{Pipeline Loss} (\%) &\textbf{Budget} (ms) & 
		\textbf{Period} (ms) 
		\\\hline
		\multirow{2}{2cm}{ProcData} & 0 & 0.2 & 2 \\\cline{2-4}
		& 20 & 0.2 & 2.5 \\\hline
		\multirow{2}{2cm}{CanWrite} & 0 & 0.1 & 2 \\\cline{2-4}
		& 20 & 0.1 & 2.5 \\\hline	
		\multirow{2}{2cm}{RTControl} & 0 & 0.1 & 2 \\\cline{2-4}
		& 20 & 0.1 & 2.5 \\\hline
	\end{tabularx}
	\caption{Task parameters for different loss constraints.}
	\label{tab:async_loss}
%        \vspace{-0.1in} 
\end{table}
\normalsize

\begin{figure}[!htb]
	\centering
	\begin{subfigure}{0.24\textwidth}
		\centering
		\includegraphics[width=\textwidth]{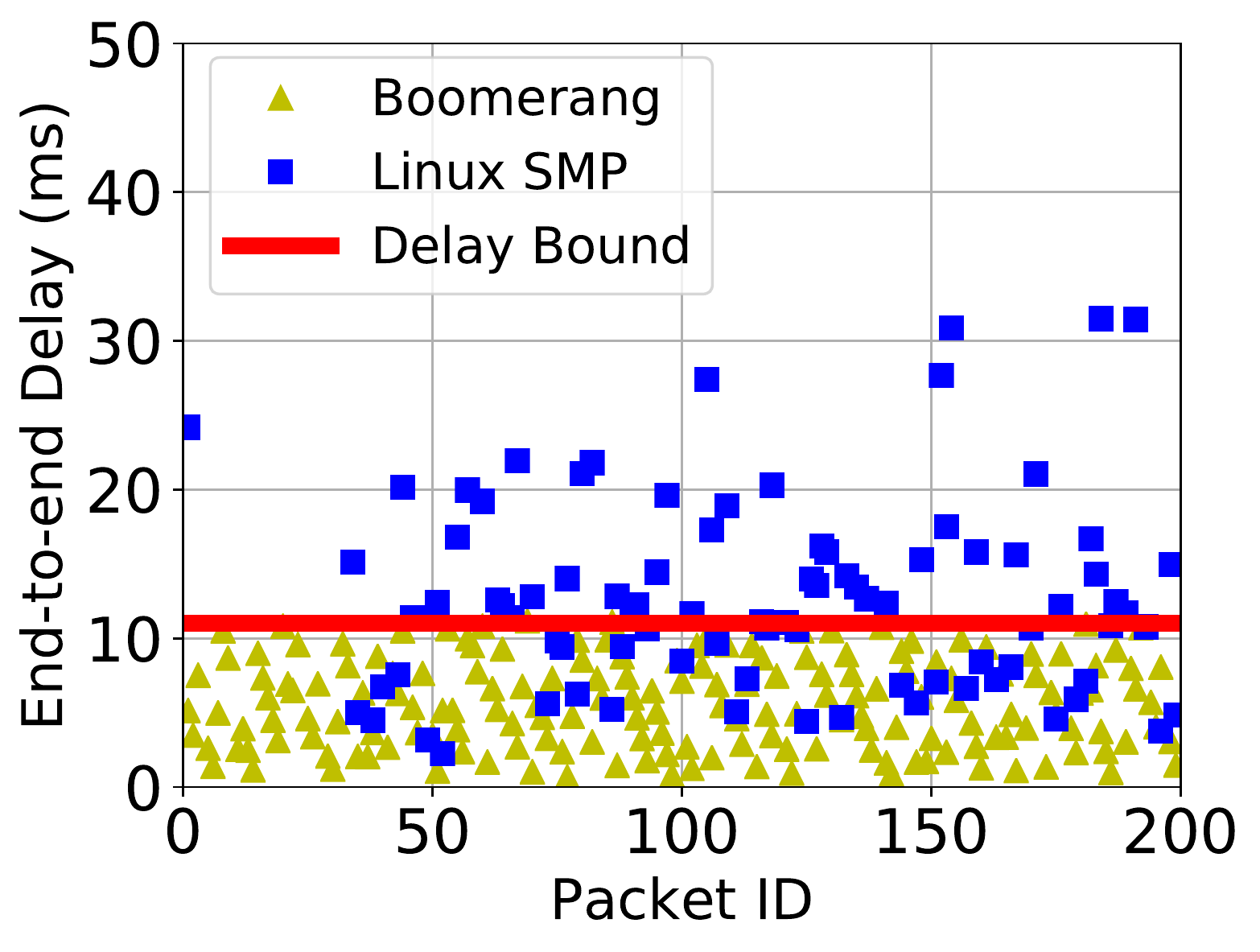}
		\caption{Pipeline 1}
		\label{fig:async_20_pipe1}
	\end{subfigure}
	\hfill
	\begin{subfigure}{0.24\textwidth}
		\centering
		\includegraphics[width=\textwidth]{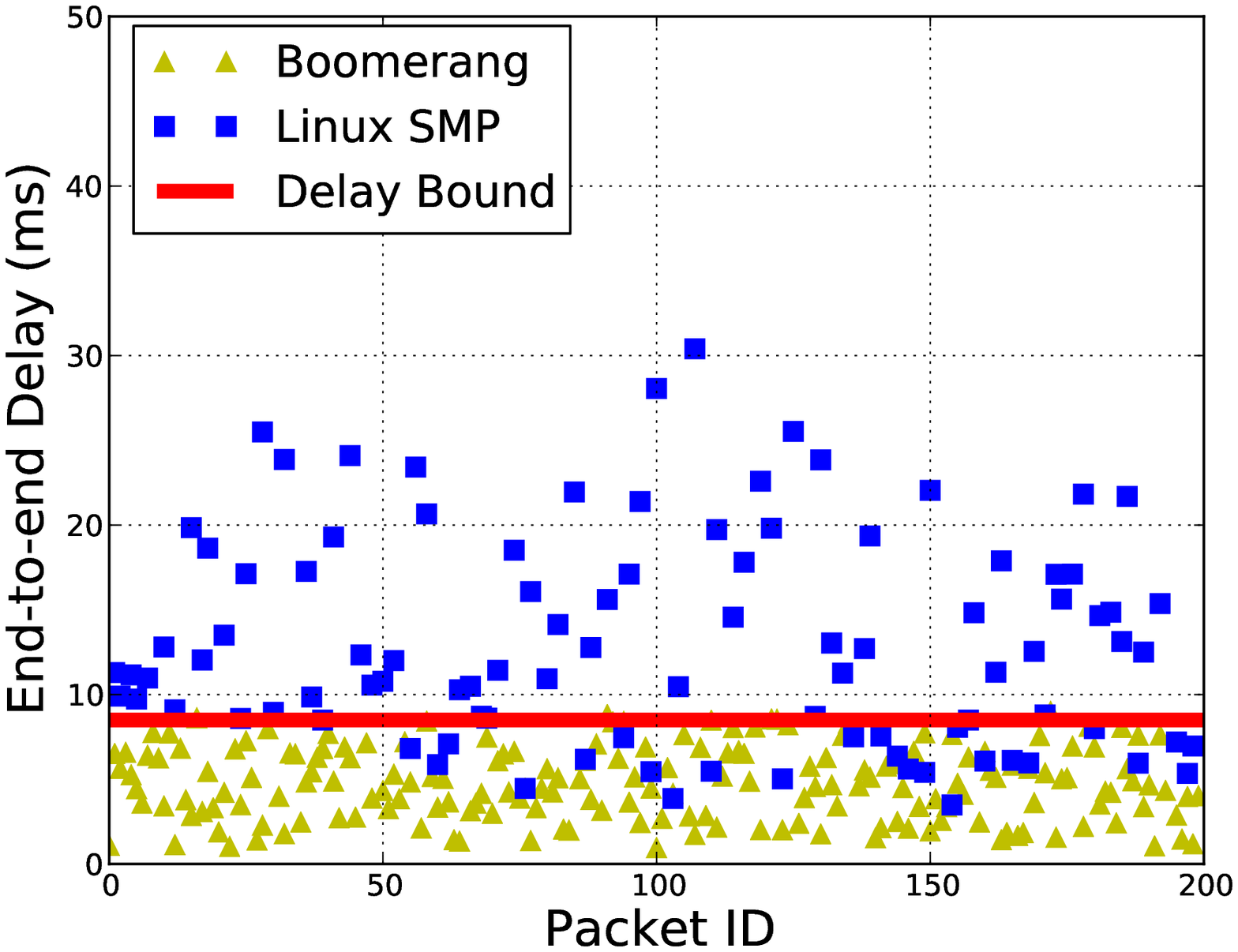}
		\caption{Pipeline 2}
		\label{fig:async_20_pipe2}
	\end{subfigure}
        \caption{End-to-end delay with 20\% allowed loss.}
%        \vspace{-0.1in}
\end{figure}

Figures~\ref{fig:async_20_pipe1} and~\ref{fig:async_20_pipe2} show the
the performance of Boomerang versus Linux SMP. Once again, both
pipelines transfer data within their end-to-end delay bounds with
Boomerang, but not with Linux SMP. The packet latency for Pipeline 2
is, on average, worse for Linux SMP in Figure~\ref{fig:async_20_pipe2}
compared to Figure~\ref{fig:async_0_pipe2}. This is because the {\tt
  RTControl} task might not receive a packet until a later period due
to lost transfers. The task period itself is also larger to cause
the increased likelihood of packet loss.

Boomerang keeps the loss ratio within 20\%, as observed in
Table~\ref{tab:async_20}. However, Linux SMP loses 50-55\% of the 200
packets sent across each pipeline.

\footnotesize
\begin{table}[!ht]
        \centering
	\begin{tabularx}{\columnwidth}{ |p{1.3cm}|X|X|X|X| }\hline
		\textbf{System} & \textbf{Min} (ms) & \textbf{Max} (ms) & 
		\textbf{Avg}  
		(ms) & 
		\textbf{Loss} (\%) \\\hline\hline
		\multicolumn{5}{|c|}{\textbf{Pipeline 1} (Delay bound = 11 ms)}\\\hline
		Boomerang & 0.64 & 10.96 & 4.87 &  3.5 \\\hline
		Linux SMP & 2.24 & 98.21 & 14.46 & 55 \\\hline\hline
		\multicolumn{5}{|c|}{\textbf{Pipeline 2} (Delay bound = 8.5 
			ms)}\\\hline
		Boomerang & 0.64 & 2.38 & 1.07 &  0 \\\hline
		Linux SMP & 3.49 & 96.02 & 13.91 & 50 \\\hline
	\end{tabularx}
	\caption{Latency - 20\% allowed loss.}
	\label{tab:async_20}
%        \vspace{-0.2in}
\end{table}
\normalsize

\subsection{ACRN Partitioning Hypervisor}
The experiments in Section~\ref{sect:async} were repeated with an
implementation of tuned pipes in the ACRN partitioning
hypervisor. ACRN has similarities to
Jailhouse~\cite{ramsauer2017look}, but is already ported to the Up
Squared board and is targeted at the same applications as Boomerang.
ACRN specifically supports safety-critical applications such as
Automotive SDC (Software Defined Cockpit) and IVE (In-Vehicle
Experience), similar to Boomerang~\cite{hotmobile2020}. It supports
partitioning of CPU cores, memory, and I/O devices among one Service
OS (SOS) and multiple User OSes (UOS). The SOS provides backend device
drivers and bootstraps UOSes.

\begin{figure}[h]
     \centering
     \includegraphics[width=\columnwidth]{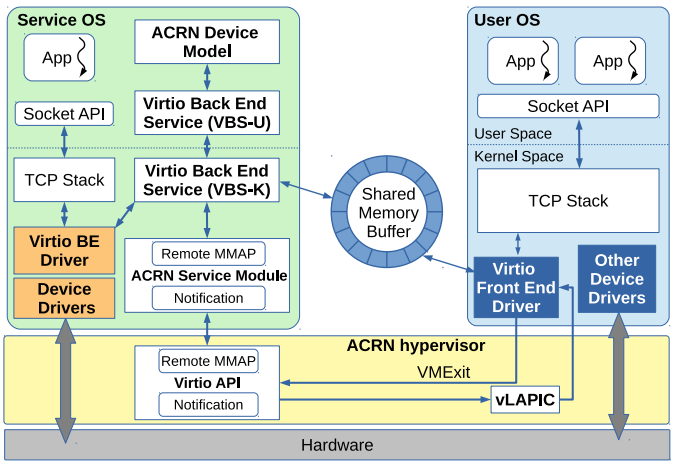}
     \caption{Inter-sandbox communication in ACRN.}
     \label{fig:acrn_isbc}
\end{figure}

\begin{figure*}[!ht]
  \begin{subfigure}{0.24\textwidth}
    \centering
    \includegraphics[width=\textwidth]{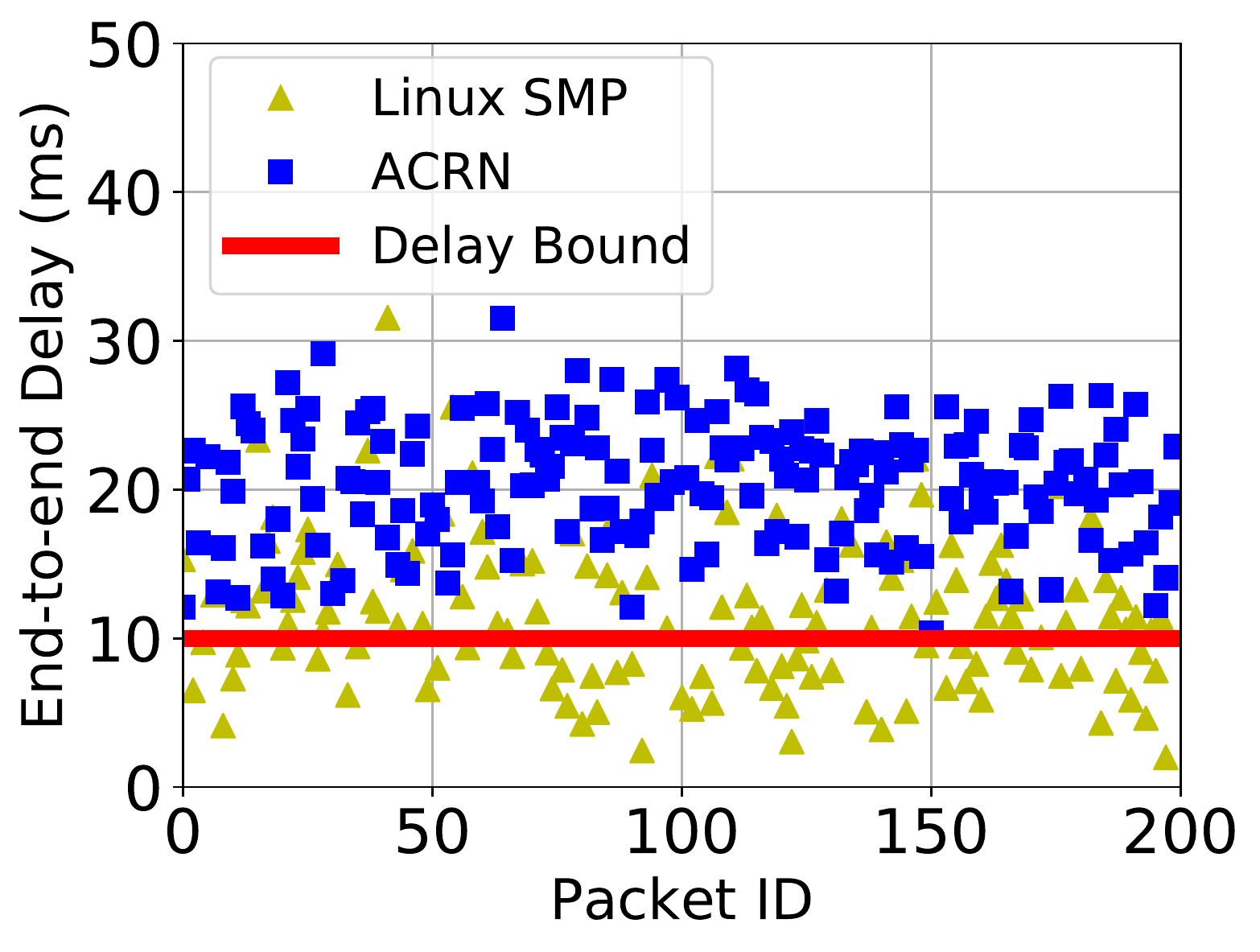}
    \caption{Pipeline 1 - no expected loss}
    \label{fig:acrn_async_0_pipe1}
  \end{subfigure}
  \begin{subfigure}{0.24\textwidth}
    \centering
	\includegraphics[width=\textwidth]{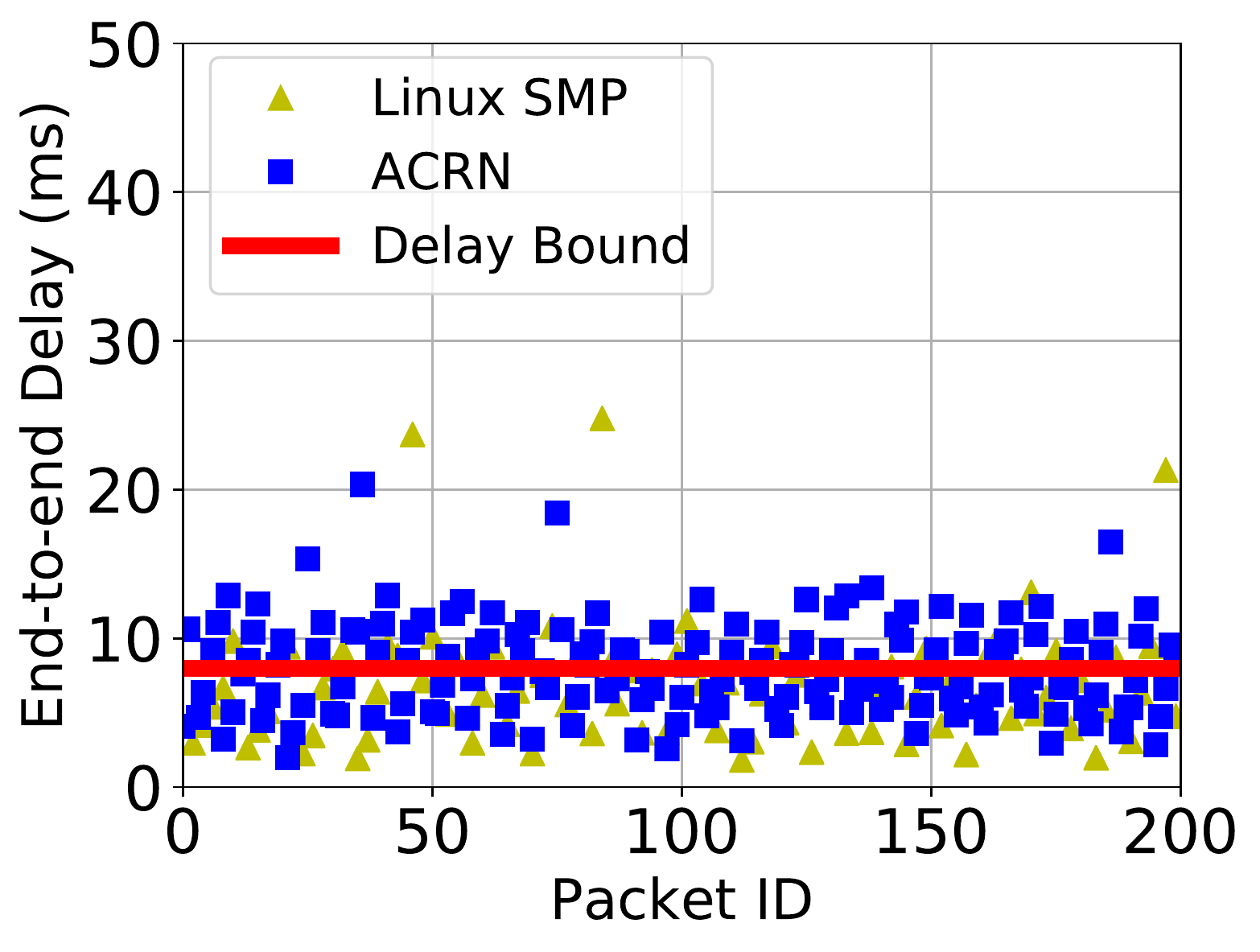}
	\caption{Pipeline 2 - no expected loss}
	\label{fig:acrn_async_0_pipe2}
  \end{subfigure}
  \begin{subfigure}{0.24\textwidth}
    \centering
    \includegraphics[width=\textwidth]{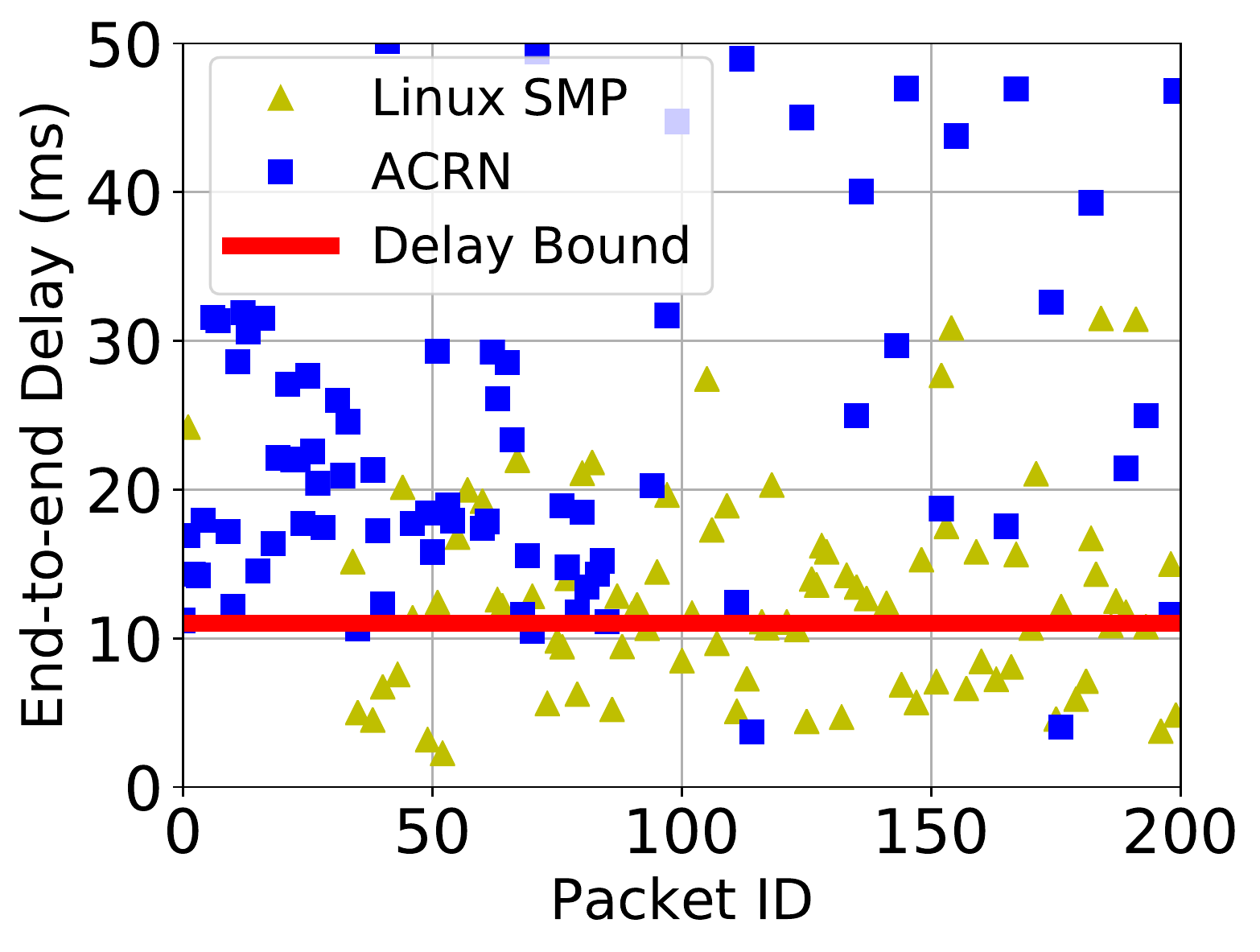}
    \caption{Pipeline 1 - 20\% allowed loss}
    \label{fig:acrn_async_20_pipe1}
  \end{subfigure}
  \begin{subfigure}{0.24\textwidth}
    \centering
    \includegraphics[width=\textwidth]{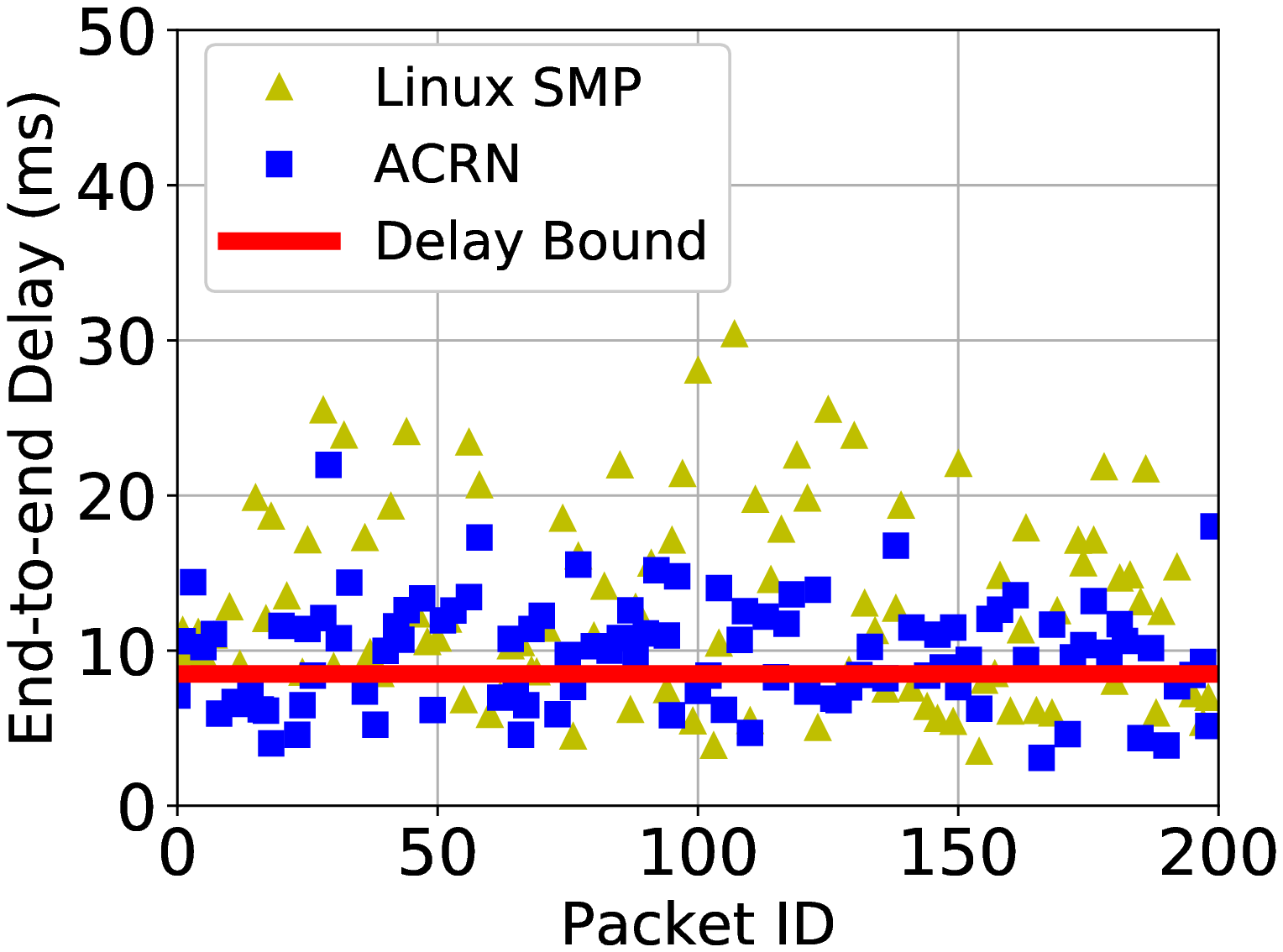}
    \caption{Pipeline 2 - 20\% allowed loss}
    \label{fig:acrn_async_20_pipe2}
  \end{subfigure}
  \caption{ACRN versus Linux SMP for asynchronous communication.}
  \vspace{-0.15in}
\end{figure*}

The ACRN tuned pipe implementation uses a virtual network bridge and
tap devices for inter-sandbox
communication. Figure~\ref{fig:acrn_isbc} depicts how data is
exchanged between a UOS and the SOS, using shared memory ring buffers
mapped to both VMs.  A UOS request passes through a TCP stack and
virtual device driver, causing a VMExit. Then the hypervisor notifies
the SOS about the new message. Virtio services within the SOS deliver
the message to the appropriate backend device driver, where it passes
through the TCP stack and into user-space. Although capable of
mimicking network communication between two guests, this approach
incurs far more timing unpredictability compared to Boomerang's
dedicated shared memory communication channels. Data exchanges between
tuned pipes in ACRN incur VMExits and, hence, control flow transitions
via the ACRN hypervisor that are avoided with Boomerang. The
consequence of this is shown in
Figures~\ref{fig:acrn_async_0_pipe1}--\ref{fig:acrn_async_20_pipe2},
where ACRN is compared with Linux SMP. Linux SMP has already been
shown to be less predictable than Boomerang in
Section~\ref{sect:async}.

In these experiments we used a PREEMPT\_RT-patched ClearOS Linux based
on kernel version 4.19.73 for both the SOS and UOS, as recommended by
the ACRN developers. Both ACRN and Linux SMP had the same mapping of
threads to cores, as shown for Boomerang in
Table~\ref{tab:pipelines_details}.  ACRN additionally partitioned
tasks and resources in the same way as Boomerang in
Figure~\ref{fig:experimental_setup}, except the SOS replaced
Boomerang's Quest RTOS, and the UOS featured ClearOS Linux instead of
Yocto Linux.  We intentionally did not port the Quest RTOS to ACRN as
it would leave little difference between the solution provided by
Boomerang and ACRN, except the implementation of the hypervisor and
inter-sandbox communication method.

As Boomerang already outperforms Linux SMP, it follows that ACRN's
lack of timing predictability makes it inferior for end-to-end
communication guarantees.

%\vspace{-0.1in}
\subsection{Synchronous Communication}
We repeated experiments with Pipelines 1 and 2 using FIFO-buffering.
The constraint solver described in Section~\ref{sect:qos} is used to
establish correct budgets, periods and buffer sizes when pipelines are
constructed. The updated budgets and periods are presented in
Table~\ref{tab:sync_pipeline}. Buffer sizes are 4, 2 and 4 messages,
respectively between {\tt CanRead} and {\tt ProcData}, {\tt ProcData}
and {\tt CanWrite}, and {\tt RTFusion} and {\tt RTControl}.

\footnotesize
\begin{table}[!ht]
	\centering
	\begin{tabularx}{\columnwidth}{ |p{1.6cm}|p{1.8cm}|p{1.75cm}|X| }\hline 
		\textbf{Task} & \textbf{Budget} (ms) & \textbf{Period} (ms) & 
		\textbf{Utilization} (\%)\\\hline\hline
		\multicolumn{4}{|c|}{\textbf{Pipeline 1} (CAN4)} \\\hline
		CanRead & 0.1 & 2 & 5 \\\hline
		ProcData & 0.2 & 4 & 5 \\\hline
		CanWrite & 0.2 & 4 & 5 \\\hline\hline
		\multicolumn{4}{|c|}{\textbf{Pipeline 2} (CAN5)} \\\hline
		RTFusion & 0.1 & 2 & 5 \\\hline
		RTControl & 0.125 & 2.5 & 5 \\\hline
	\end{tabularx}
	\caption{Synchronous pipeline \emph{(common threads not shown).}}
	\label{tab:sync_pipeline}
\end{table}
%\vspace{-0.2in}
\normalsize

%\vspace{-0.1in}
\subsubsection{Throughput and Delay}

\begin{figure}[!ht]
	\centering
	\begin{subfigure}{0.24\textwidth}
		\includegraphics[width=\textwidth]{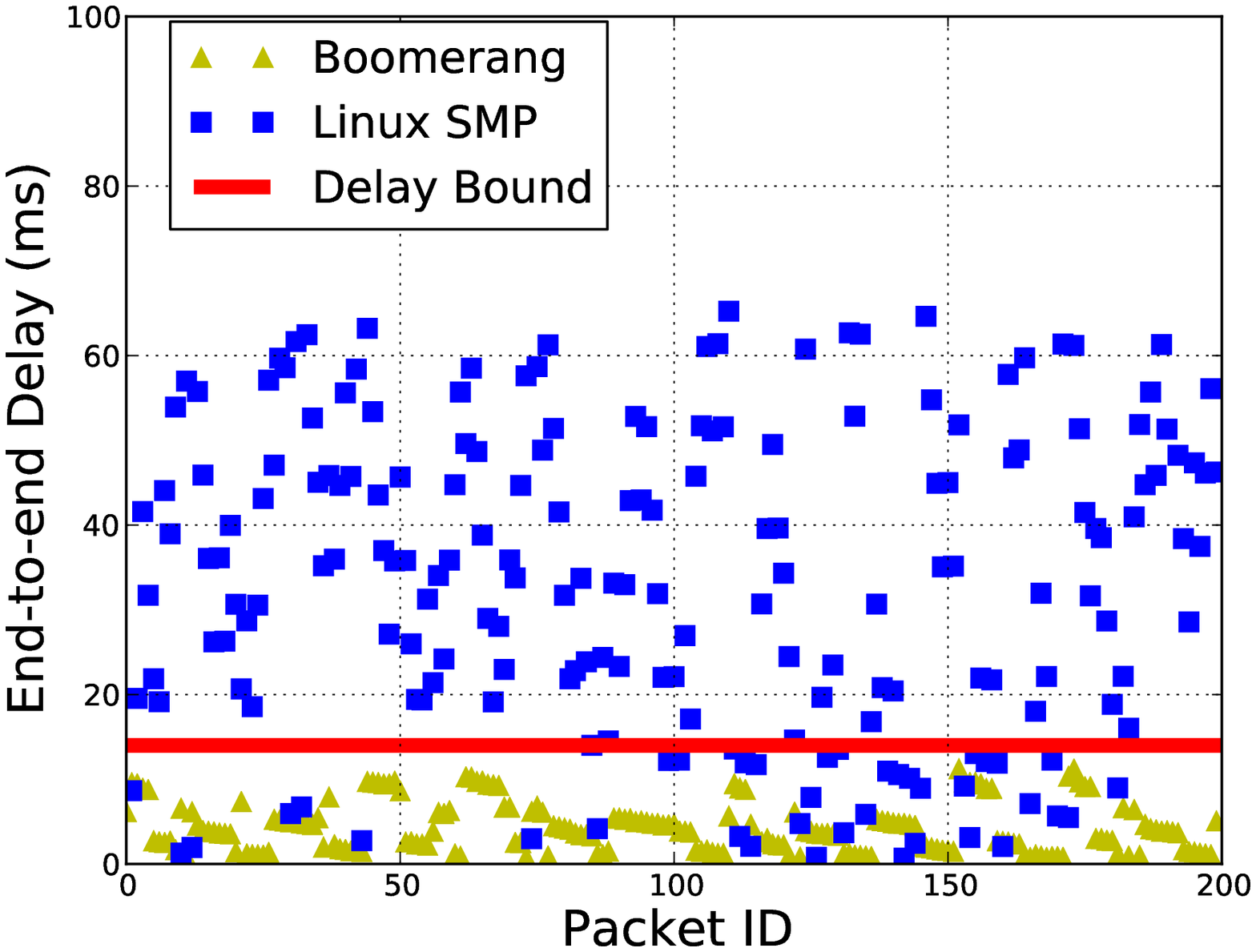}
		\caption{Pipeline 1}
		\label{fig:fifo_latency_pipe1}
	\end{subfigure}
	\begin{subfigure}{0.24\textwidth}
		\includegraphics[width=\textwidth]{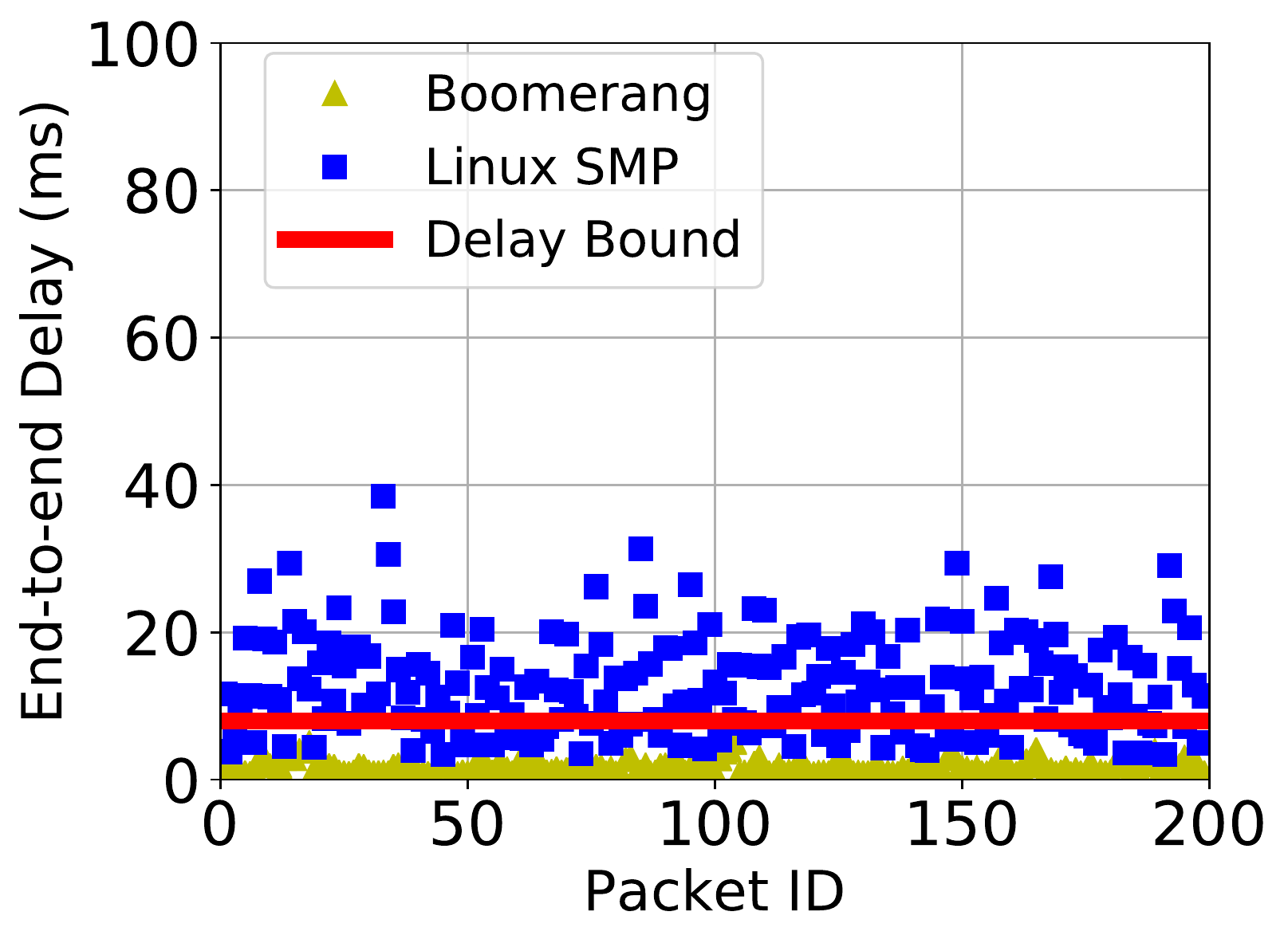}
		\caption{Pipeline 2}
		\label{fig:fifo_latency_pipe2}
	\end{subfigure}
        \caption{FIFO buffered synchronous communication.}
%        \vspace{-0.2in}
\end{figure}

The expected end-to-end delay of Pipeline 1 is increased to 14ms because of
the increased periods of the \texttt{tpipe}
threads. Figures~\ref{fig:fifo_latency_pipe1} and
~\ref{fig:fifo_latency_pipe2} show the revised end-to-end delays. Measurements
are summarized in Table~\ref{tab:sync_latency}. FIFO buffering does not
improve the latency for Linux SMP because of previously mentioned issues with
interrupts. However, it reduces the packet loss for Linux SMP, as a
buffer holds messages even if a {\tt tpipe} thread is interrupted.

\footnotesize
\begin{table}[!ht]
	\begin{tabularx}{\columnwidth}{ |p{1.3cm}|X|X|X|X| }\hline
		\textbf{System} & \textbf{Min} (ms) & \textbf{Max} (ms) & 
		\textbf{Avg}  
		(ms) & 
		\textbf{Loss} (\%) \\\hline\hline
		\multicolumn{5}{|c|}{\textbf{Pipeline 1} (Delay bound = 14 ms)}\\\hline
		Boomerang & 0.77 & 11.23 & 4.25 &  0 \\\hline
		Linux SMP & 0.96 & 65.24 & 33.10 & 0.5 \\\hline\hline
		\multicolumn{5}{|c|}{\textbf{Pipeline 2} (Delay bound = 8.5 
			ms)}\\\hline
		Boomerang & 0.70 & 5.03 & 1.56 &  0 \\\hline
		Linux SMP & 0.70 & 38.46 & 12.84 & 0 \\\hline
	\end{tabularx}
	\caption{Latency - FIFO buffering.}
	\label{tab:sync_latency}
\end{table}
\normalsize

Table~\ref{tab:sync_throughput} shows the throughput with Boomerang and Linux
SMP are similar, although the standard deviation is smaller with
Boomerang. Arrival rates ($\lambda$) from CAN4 and CAN5 are shown for each
pipeline.

\footnotesize
\begin{table}[!ht]
	\begin{tabularx}{\columnwidth}{ |p{1.3cm}|X|X|X|p{0.7cm}| }\hline
		\textbf{System} & \textbf{Min} (msg/s) & \textbf{Max} (msg/s) & 
		\textbf{Avg} (msg/s) & 
		\textbf{stddev} \\\hline\hline
		\multicolumn{5}{|c|}{\textbf{Pipeline 1} ($\lambda$=100 msgs/s)}\\\hline
		Boomerang & 99 & 101 & 99.77 & 0.63 \\\hline
		Linux SMP & 86 & 105 & 98.1 & 4.39 \\\hline\hline
		\multicolumn{5}{|c|}{\textbf{Pipeline 2} ($\lambda$=125 msgs/s)}\\\hline
		Boomerang & 123 & 126 & 124.77 & 0.73 \\\hline
		Linux SMP & 120 & 126 & 123.17 & 1.39 \\\hline
	\end{tabularx}
	\caption{Synchronous throughput.}
	\label{tab:sync_throughput}
%        \vspace{-0.15in}
\end{table}
\normalsize

\subsection{MIMO Pipelines}
Boomerang supports the construction of pipelines with multiple inputs
and outputs (MIMO). We constructed a pipeline based on
Figure~\ref{fig:pipeline_example}, representative of automotive tasks
where multiple sensor inputs are combined to control more than one
actuator. Using the labeling in that figure, tuned pipes $A-F$ have
the following (budget, period) pairs in milliseconds: $A$~(0.1,1),
$B$~(0.2, 2), $C$~(0.1, 1), $D$~(0.4, 2), $E$~(0.1, 1) and $F$~(0.1,
1). $A$ reads input from CAN4 while $C$ reads input from
CAN5. Similarly, $E$ writes back to CAN4, and $F$ writes to CAN5. The
CAN4 path traverses $ABDE$, while the CAN5 path traverses $CDF$. Tuned
pipe $D$ is shared by both paths; it runs in Linux while all other
tuned pipes operate in the RTOS.

\footnotesize
\begin{table}[!ht]
	\begin{subfigure}{0.5\columnwidth}
		\begin{tabularx}{\textwidth}{ |X|X|X|X|X| }\hline
			\textbf{Min} (ms) & \textbf{Max} (ms) & 
			\textbf{Avg}  
			(ms) & \textbf{Std Dev} &
			\textbf{Loss} (\%) \\\hline\hline
			\multicolumn{5}{|c|}{\textbf{CAN4 path} (Delay bound=10 ms)}\\\hline
			0.86 & 9.63 & 2.56 & 1.32 & 0 \\\hline\hline
			\multicolumn{5}{|c|}{\textbf{CAN5 path} (Delay bound=8 ms)}\\\hline
			0.70 & 5.00 & 2.11 & 0.86 & 0 \\\hline
		\end{tabularx}
		\caption{Latency}
		\label{tab:mimo_latency}
	\end{subfigure}\hfill
	\begin{subfigure}{0.5\columnwidth}
		\begin{tabularx}{\textwidth}{ |X|X|X|X| }\hline
			\textbf{Min} (msg/s) & \textbf{Max} (msg/s) & 
			\textbf{Avg} (msg/s) & 
			\textbf{Std Dev} \\\hline\hline
			\multicolumn{4}{|c|}{\textbf{~CAN4 path~} ($\lambda$=100 msgs/s)}\\\hline
			99 & 101 & 99.74 & 0.58 \\\hline\hline
			\multicolumn{4}{|c|}{\textbf{~CAN5 path~} ($\lambda$=125 msgs/s)}\\\hline
			124 & 126 & 124.74 & 0.58 \\\hline
		\end{tabularx}
		\caption{Throughput}
		\label{tab:mimo_throughput}
	\end{subfigure}
        \caption{MIMO pipelines in Boomerang.}
\end{table}
\normalsize

\begin{figure}[!ht]
        \centering
%        \vspace{-0.1in}
 	\includegraphics[width=0.5\columnwidth]{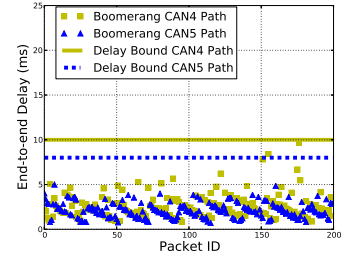}
%        \vspace{-0.05in}
 	\caption{MIMO pipeline delay guarantees.}
 	\label{fig:mimo_latency}
%        \vspace{-0.2in}
\end{figure}

Tables~\ref{tab:mimo_latency} and ~\ref{tab:mimo_throughput} summarize
the latencies and throughput, while Figure~\ref{fig:mimo_latency}
shows the end-to-end delay. The delay bounds of the two paths are 10ms
and 8ms, accounting for 4ms worst-case delay from {\em mhydra rx/tx} and {\em USB\_BH} threads, using the parameters shown in
Table~\ref{tab:pipelines_details}. Even with multiple device inputs
and outputs, both paths through CAN4 and CAN5 transfer data within
their expected bounds.

\section{Related Work}
\label{sect:related}
\subsection{Operating Systems}
Mercer et al implemented {\em processor capacity reserves} in the Mach
micro-kernel~\cite{Mercer94processorcapacity}, to provide tasks with
budgets and periods. Steere et al used a reservation-based scheme
along with a feedback-based controller to adjust CPU allocations among
tasks~\cite{Steere_Goel_Gruenberg_McNamee_Pu_Walpole_1999}.  Linux
supports reservation-based scheduling using the \texttt{PREEMPT\_RT}
patch~\cite{preemptrt} and \texttt{SCHED\_DEADLINE} \cite{scheddll}
task execution managed by a Constant Bandwidth
Server~\cite{cbs}. LITMUS\textsuperscript{RT}
\cite{calandrino2006litmus} is a Linux-based system that supports
configurable real-time schedulers, including those with
reservations. Multiple RTOSes attempt to provide temporal isolation to
tasks~\cite{vxworks, mert, microc-osii}. However, these systems do not
properly handle events such as interrupts, which may interfere with
the timing requirements of real-time tasks.

%% Boomerang extends this method of temporal
%% isolation to encompass I/O event handling (i.e., bottom half interrupt
%% processing).

RT-Linux virtualizes interrupts for non-time-critical parts of the
system, thereby ensuring real-time service to time-critical
tasks~\cite{rtlinux}. Similar approaches have been adopted by Wind
River Linux~\cite{wind-river-linux}, the Real-time Application
Interface (RTAI) for Linux~\cite{dozio2003linux}, Xenomai
\cite{xenomai}, and NASA's CFS Linux~\cite{cfslinux}.  Zhang et al
integrated interrupt handling with task scheduling in Linux. A bottom
half handler for a device interrupt inherited the highest priority of
a blocked process waiting on the device~\cite{process-aware}. However,
interrupt handling was not limited to a CPU reservation, meaning a
burst of interrupts could still interfere with tasks.

Many real-time OSes provide a single address space, multi-threaded
solution for multicore machines~\cite{ecos,rtems,freertos}. However,
this is insufficient for many safety-critical domains, requiring both
temporal and spatial isolation between components of different
criticality levels.  The Quest RTOS~\cite{quest} not only supports
multiple address spaces, but also provides a Priority-Inherited
Bandwidth-preserving Server approach to serve the interrupts in a
timely manner along with CPU-bound tasks. While Quest provides timing
isolation for both I/O- and CPU-bound tasks, it does not support the
richness of services found in a legacy system such as Linux.
%\vspace{-0.1in}
\subsection{Hypervisors}
Several hypervisors attempt to support both temporal and spatial
isolation of guests~\cite{windriverhypervisor, mentorhypervisor,
  evm,masmano2009xtratum}. RT-Xen~\cite{xi2011rt} adds real-time
scheduling support to the Xen~\cite{barham2003xen}
hypervisor. However, all these hypervisors {\em multiplex} their
guests on a shared physical machine. They virtualize interrupts, and
perform additional resource management operations that conflict with
the policies within each guest.

Partitioning hypervisors allow guests to directly manage subsets of
machine resources.
%The hypervisor is removed from I/O and resource
%management once each guest is granted access to its machine resources.
The Quest-V \cite{quest-v-toc} separation
kernel~\cite{Rushby81:SOSP} uses a partitioning hypervisor to
support the co-existence of the Quest RTOS and one or more general
purpose OSes. Each guest OS runs simultaneously on separate cores in a
multicore machine, with device interrupts delivered directly to the
guest that owns the device.

PikeOS~\cite{pikeos} and Muen~\cite{buerki2013muen} are separation kernels
that support multiple guest OSes. However, unlike Quest-V, interrupts are
trapped into the hypervisor, and subsequently delivered to the guest
OSes. Jailhouse~\cite{ramsauer2017look} and ACRN~\cite{acrn} have similarities
to Quest-V. Jailhouse uses Linux to bootstrap a system that provides cells for
system inmates. These are essentially restricted hardware subsets assigned to
guests.
%However, Jailhouse does not currently provide an integrated approach
%for guests to communicate in real-time via shared memory channels.
ACRN's
philosophy is to allow a service OS to manage machine resources on behalf of
other safety-critical OSes. However, as with Jailhouse, there is currently no
way to communicate between guests with end-to-end timing
guarantees. Boomerang's partitioning hypervisor is modeled on the approach
taken by Quest-V, but provides support for composable tuned pipes spanning
multiple guests.

%\vspace{-0.1in}
\subsection{Predictable Communication}
Boomerang's support for composable tuned pipes is inspired by
Scout~\cite{scout}, which treats {\em paths} through a sequence of
services as first-class schedulable entities. Path processing is
entirely within the context of a single thread that is scheduled
according to the bottleneck queue. Boomerang, in contrast, schedules
each component of a pipeline with a separate time-budgeted
thread. This allows paths to be interleaved and executed on different
PCPUs, spanning different sandboxes.

RAD-FLOWS~\cite{pineiro2011rad} provided a design framework for
predictable data communication.
Golchin et al developed a system abstraction for predictable data delivery
between USB devices and a real-time
process~\cite{Golchin_Cheng_West_2018}. Boomerang provides support for
real-time I/O to span multiple tasks in different guest VMs.

\section{Conclusions and Future Work}
\label{sect:conc}
This paper presents Boomerang, an I/O system comprising real-time task
pipelines in a partitioning hypervisor. Boomerang's partitioning
hypervisor connects a built-in guest RTOS (Quest) with a legacy
system such as Linux, via secure and predictable shared memory
communication channels.
The legacy OS benefits from timing predictable services that are isolated from
less critical code. At the same time, the RTOS benefits from the pre-existing
services, including libraries and lower criticality device drivers of a legacy
non-real-time system.

Boomerang supports composable tuned pipes, for real-time task pipelines that
require guaranteed end-to-end service on data transfers. 
The system provides real-time task pipelines with complementary
legacy services that are timing predictable using CPU reservations.

Experiments show that real-time task pipelines guarantee end-to-end
throughput, delay and loss requirements in Boomerang. This is the case
for all pipelines contained within the RTOS and which span both the
RTOS and Linux. In contrast, task pipelines in a Linux-only system are
not able to ensure end-to-end service constraints, even when using CPU
reservations. This is because of task interference by interrupts from
I/O devices. The interrupt handlers need to be assigned suitable CPU
reservations at appropriate priorities that match the pipelined tasks
they serve. Alternatively, if I/O processing is assigned to a
dedicated core, it reduces system utilization. Finally, other
partitioning hypervisors such as ACRN rely on heavyweight networking
protocols and VMExits to perform inter-guest communication via shared
memory, rendering them unsuitable for real-time data processing.

Future work will extend Boomerang's composable tuned pipes to span different
physical machines. We see a programming model for real-time pipes useful in
data flow machines and stream processing applications, such as those in
neuromorphic computing.
 
\section*{Acknowledgment}
This work is supported in part by the National Science Foundation
(NSF) under Grant \# 1527050.  Any opinions, findings, and conclusions
or recommendations expressed in this material are those of the
author(s) and do not necessarily reflect the views of the NSF. Thanks
also to Drako Motors for equipment used in this work. This paper has
been greatly improved by the feedback of the RTAS reviewers.

Source code for this work will be made available via www.questos.org.


\end{document}